\documentclass[twocolumn]{aastex631}
\usepackage{graphicx}
\usepackage{comment}
\usepackage{xspace}
\usepackage{natbib}
\usepackage{amsmath}
\usepackage{graphics}
\usepackage{array}
\newcolumntype{L}[1]{>{\raggedright\arraybackslash}p{#1}}
\usepackage{tabularx}
\usepackage{CJKutf8}            % Chinese name

\newcommand{\noop}[1]{}

\newcommand{\CIV}{\ion{C}{4}}
\newcommand{\lya}{Ly$\alpha$\xspace}
\newcommand{\nlaefull}{79,830}
\newcommand{\nlaegood}{70,691}
\newcommand{\nlan}{33,612}

\newcommand{\labdensity}{420}
\newcommand{\zhet}{\texttt{z\_hetdex}}

\newcommand{\hetg}{$g_\mathrm{HETDEX}$}
\newcommand{\fluxunit}{$10^{-17}$~erg~s$^{-1}$~cm$^{-2}$}

\newcommand{\sbunits}{$10^{-18}$~erg~s$^{-1}$~cm$^{-2}~\mathrm{arcsec}^{-2}$}
\newcommand{\sbunitsraw}{erg~s$^{-1}$~cm$^{-2}~\mathrm{arcsec}^{-2}$\xspace}

\begin{document}

\title{\lya\ Nebulae in HETDEX:\\
The Largest Statistical Census Bridging Ly$\alpha$ Halos and Blobs across Cosmic Noon}

\author[0000-0002-2307-0146]{Erin Mentuch Cooper}
\affiliation{Department of Astronomy, The University of Texas at Austin, 2515 Speedway Boulevard, Austin, TX 78712, USA}
\email{erin.hetdex@gmail.com}

\author[0000-0002-8433-8185]{Karl Gebhardt}
\affiliation{Department of Astronomy, The University of Texas at Austin, 2515 Speedway Boulevard, Austin, TX 78712, USA}

\author[0000-0002-8925-9769]{Dustin Davis}
\affiliation{Department of Astronomy, The University of Texas at Austin, 2515 Speedway Boulevard, Austin, TX 78712, USA}

\author[0000-0002-1328-0211]{Robin Ciardullo}
\affiliation{Department of Astronomy \& Astrophysics, The Pennsylvania State University, University Park, PA 16802, USA}
\affiliation{Institute for Gravitation and the Cosmos, The Pennsylvania State University, University Park, PA 16802, USA}

\author[0000-0002-0885-8090]{Chris Byrohl}
%\affiliation{Kavli Institute for the Physics and Mathematics of the Universe (WPI), Todai Institutes for Advanced Study, the University of Tokyo, Kashiwanoha, Kashiwa, Chiba 277-8583, Japan}
\affiliation{Kavli Institute for the Physics and Mathematics of the Universe (WPI), University of Tokyo, Kashiwa, Chiba 277-8583, Japan}
\affiliation{Institut f\"{u}r Theoretische Astrophysik, ZAH, Universit\"{a}t Heidelberg,  Albert-Ueberle-Str. 2, 69120 Heidelberg, Germany}

\author[0000-0001-5561-2010]{Chenxu Liu\begin{CJK*}{UTF8}{gkai} (刘辰旭) \end{CJK*}}
\affiliation{South-Western Institute for Astronomy Research, Key Laboratory of Survey Science of Yunnan Province, Yunnan University, Kunming, Yunnan 650500, People's Republic of China}

\author[0000-0002-1998-5677]{Maya H. Debski}
\affiliation{Department of Astronomy, The University of Texas at Austin, 2515 Speedway Boulevard, Austin, TX 78712, USA}

\author[0000-0003-2332-5505]{\'{O}scar A. Ch\'{a}vez Ortiz}
\affiliation{Department of Astronomy, The University of Texas at Austin, 2515 Speedway Boulevard, Austin, TX 78712, USA}

\author[0000-0002-7025-6058]{Maximilian Fabricius}
\affiliation{Max Planck Institute for Extraterrestrial Physics, Giessenbachstr. 1, 85748 Garching, Germany}
\affiliation{Universit\"ats-Sternwarte M\"unchen, Fakult\"at f\"ur Physik, Ludwig-Maximilians-Universit\"at M\"unchen, Scheinerstrasse 1, 81679 M\"unchen, Germany}

\author[0000-0003-2575-0652]{Daniel J. Farrow}
\affiliation{Centre of Excellence for AI, Data Science and Modelling (DAIM), University of Hull, Cottingham Road, Hull, HU6 7RX, UK}
\affiliation{E. A. Milne Centre for Astrophysics, University of Hull, Cottingham Road, Hull, HU6 7RX, UK}

\author[0000-0001-8519-1130]{Steven L. Finkelstein}
\affiliation{Department of Astronomy, The University of Texas at Austin, 2515 Speedway Boulevard, Austin, TX 78712, USA}

\author[0000-0001-6842-2371]{Caryl Gronwall}
\affiliation{Department of Astronomy \& Astrophysics, The Pennsylvania
State University, University Park, PA 16802, USA}
\affiliation{Institute for Gravitation and the Cosmos, The Pennsylvania State University, University Park, PA 16802, USA}

\author[0000-0001-6717-7685]{Gary J. Hill} 
\affiliation{Department of Astronomy, The University of Texas at Austin, 2515 Speedway Boulevard, Austin, TX 78712, USA} 
\affiliation{McDonald Observatory, The University of Texas at Austin, 2515 Speedway Boulevard, Austin, TX 78712, USA}

\author[0000-0002-6907-8370]{Maja Lujan Niemeyer}
\affiliation{Max-Planck-Institut f\"{u}r Astrophysik, Karl-Schwarzschild-Str. 1, 85741 Garching, Germany}

\author[0000-0002-8468-9575]{Brianna McKay}
\affiliation{Department of Astronomy, University of Washington, Seattle, 3910 15th Ave NE, Room C319, Seattle WA 98195-0002}

\author[0000-0003-3823-8279]{Shiro Mukae}
\affiliation{Department of Astronomy, The University of Texas at Austin, 2515 Speedway Boulevard, Austin, TX 78712, USA}
\affiliation{MIRAI Technology Institute, Shiseido Co., Ltd., 1-2-11, Takashima, Nishi-ku, Yokohama, Kanagawa, 222-0011, Japan}

\author[0000-0002-1049-6658]{Masami Ouchi}
\affiliation{National Astronomical Observatory of Japan, 2-21-1 Osawa, Mitaka, Tokyo 181-8588, Japan}
\affiliation{Institute for Cosmic Ray Research, The University of Tokyo, 5-1-5 Kashiwanoha, Kashiwa, Chiba 277-8582, Japan}
\affiliation{Department of Astronomical Science, SOKENDAI (The Graduate University for Advanced Studies), Osawa 2-21-1, Mitaka, Tokyo, 181-8588, Japan}
\affiliation{Kavli Institute for the Physics and Mathematics of the Universe (WPI), University of Tokyo, Kashiwa, Chiba 277-8583, Japan}

\author[0000-0001-8887-2257]{Huub Röttgering}
\affiliation{Leiden Observatory, Leiden University, P.O. Box 9513, 2300 RA Leiden, The Netherlands}

\author[0000-0001-7240-7449]{Donald P. Schneider}
\affiliation{Department of Astronomy \& Astrophysics, The Pennsylvania State University, University Park, PA 16802, USA}
\affiliation{Institute for Gravitation and the Cosmos, The Pennsylvania State University, University Park, PA 16802, USA}

\author[0000-0002-7327-565X]{Sarah Tuttle}
\affiliation{Department of Astronomy, University of Washington, Seattle, 3910 15th Ave NE, Room C319, Seattle WA 98195-0002}

\author[0000-0003-2977-423X]{Lutz Wisotzki}
\affiliation{Leibniz-Institut for Astrophysik Potsdam (AIP), An der Sternwarte 16, 14482 Potsdam, Germany}

\author[0000-0003-2307-0629]{Gregory Zeimann}
\affiliation{Hobby Eberly Telescope, University of Texas, Austin, Austin, TX, 78712}

\author[0000-0001-6677-9940]{Sai Zhai}
\affiliation{Leiden Observatory, Leiden University, P.O. Box 9513, 2300 RA Leiden, The Netherlands}

%\end{comment}

\begin{abstract}

The Hobby-Eberly Dark Energy Experiment (HETDEX) is an untargeted $\sim$540\,deg$^2$ spectroscopic survey of \lya emission in the  $1.9<z<3.5$ Universe.  In surface brightness, this survey reaches $1\sigma$ \lya sensitivities of $\sim2-5~\times~$\sbunits, allowing large samples of extended Ly$\alpha$ nebulae (LAN) to be studied. We selected a sample of \nlaegood\ \lya-emitting galaxies (LAEs) with an emission-line signal-to-noise ratio greater than 6 and modeled the \lya\ emission as a point-source component with an optional exponential envelope. Half ($\sim47.5\%$) of the LAE sample (\nlan\ objects) exhibits significant extended emission and is best fit by the two-component model. The fraction of resolved sources increases with Ly$\alpha$ flux and luminosity. Their isophotal areas range from 10–130\,arcsec$^2$ (median 15\,arcsec$^2$), with integrated Ly$\alpha$ fluxes from $6-2000~\times$ \fluxunit\ (median 20 × \fluxunit). Comparison between point-spread-function-weighted and isophotal flux measurements shows that the HETDEX pipeline underestimates the total \lya\ flux by $\sim$30\% on average, reflecting the substantial halo contribution in extended sources. Approximately \labdensity\ LANs are found per deg$^2$ over 79.5\,deg$^2$ of non-contiguous sky. About 12\% of resolved sources show active galactic nuclei signatures and are bright in Ly$\alpha$ and continuum. The remaining 88\% span a wide range of morphologies and often lack continuum counterparts. Exponential scale lengths show no strong correlation with Ly$\alpha$ flux or luminosity (median $11.6\pm1.9$\,kpc). Only 2.9\% of the full S/N$>6$ LAE population with ancillary data have radio counterparts, but 64\% of those are found to be extended, with the radio fraction increasing with Ly$\alpha$ size. We present a catalog of all modeled sources, with their positions, redshifts, luminosities, and structural parameters for over 70,000 LAEs consisting of 33,000 spatially extended Ly$\alpha$ nebulae. The catalog can be found at \url{https://hetdex.org/data-results/} and in the online version of this paper.

\end{abstract}


\section{Introduction} 
\label{sec:intro}

\lya\ emission is one of the most prominent spectral features in the high-redshift Universe, observable from the ground as the UV \lya\ line is redshifted into the optical. The line’s strength enables the detection of galaxies that are too faint to be seen in broadband continuum surveys, as ionizing photons from young stars and active nuclei are reprocessed into Ly$\alpha$ emission through hydrogen recombination. Not only does \lya trace stellar and active galactic nuclei (AGN) light within galaxies, it can also reveal ionized and resonantly scattered emission in the surrounding circumgalactic medium.

Spatially extended \lya emission is observed across a wide range of physical scales and surface brightnesses. \lya\ emitters (LAEs), galaxies detected through their strong \lya emission, are often accompanied by faint \lya\ halos (LAHs) on scales up to tens of proper kiloparsecs, kpc, observed across redshifts at $\sim 10^{-19}$\,\sbunitsraw sensitivities \citep{steidel2011, Hayes2014, Wisotzki2016, Leclercq2017, LujanNiemeyer22a}. LAHs might not only be a common feature of LAEs, but also ubiquitous for star-forming galaxies \citep{LujanNiemeyer22}.  For massive systems, bright extended Ly$\alpha$ structures are detected on scales ranging from tens to hundreds of kiloparsecs—and in some cases approaching megaparsec scales—at surface-brightness levels of $\sim10^{-18}$–$10^{-20}$\,\sbunitsraw, revealing filamentary gas associated with \lya\ blobs (LABs) and the cosmic web \citep[e.g.,][]{matsuda11, cantalupo14, hennawi15, borisova16, cai17, Bacon21, Martin23, tornotti2025}.

A number of physical processes can source and give rise to extended \lya structures, which we will collectively refer to as Lyman Alpha Nebulae (LANs) throughout this paper encompassing both LAH and LAB nomenclature. Central star-forming galaxies and AGN produce a large number of \lya photons following recombinations in their immediate vicinity. Subsequent resonant scatterings can illuminate the diffuse surroundings of these systems~\citep{Zheng2010,Byrohl2021,Byrohl2023}. Satellite galaxies can provide an additional \lya source, directly injected within the halo surroundings~\citep{Mas-Ribas17}. In situ \lya emission through recombinations and collisional excitations from the diffuse gas itself can be major contributors. Recombinations can be powered by escaping ionizing photons from the host halo galaxies and AGNs, as well as from the metagalactic ultraviolet background~\citep{Cantalupo2005,Kollmeier2010}. Gravitational cooling, particularly through filamentary gas streams into the dark matter halos~\citep{Dekel09}, can be significant energy source with up to $50\%$ of the cooling budget emitted through \lya~\citep{Fardal2001,Dijkstra2009}. Likely different physical processes contribute with different relative importance in different halo mass regimes. Large observational samples and their statistical analysis will disentangle these scenarios and their contributions.

Previous surveys have revealed populations of Ly$\alpha$ halos and nebulae using two complementary approaches. Narrowband imaging with Subaru/HSC, Very Large Telescope (VLT)/VIMOS, and DECam has identified thousands of emitters at discrete redshift slices (\citealt{Ouchi2008, Sobral2018, Li2024}), including $\sim$300–500 extended \lya\ nebulae in wide-area programs such as MAMMOTH-Subaru \citep{Li2024}. The MAMMOTH-Subaru survey established the first statistical view of hundreds of LANs, showing that only a small fraction are AGN or radio associated while most ($\sim$80\%) arise around UV-faint, likely dusty star-forming galaxies that may host obscured AGN \citep{Li2024}. While these studies are pioneering in scale, narrowband imaging is limited to discrete redshift windows and relies on photometric emission-line selection.

Integral-field spectroscopy with instruments such as Multi-Unit Spectroscopic Explorer (MUSE; \citealt{Bacon2015}) and Keck Cosmic Web Imager (KCWI; \citealt{kcwi2018}) enables three-dimensional mapping of extended \lya structures with exquisite sensitivity, but current surveys only cover preselected targets over small solid angles based on deep photometric imaging as is done, for example, in MUSE \citep{Wisotzki2016, Leclercq2017, Kusakabe2022} and KCWI \citep{chen2021}. These targeted \lya studies have revealed the detailed morphologies and kinematics of \lya\ halos but lack the statistical power to probe rare luminous systems or mitigate cosmic variance.

The Hobby-Eberly Telescope Dark Energy Experiment \citep[HETDEX;][]{Gebhardt2021} provides a unique statistical window into extended \lya structures. Its wide-area, random-tiling, integral-field spectroscopic survey enables the first truly volumetric census of \lya-emitting galaxies and their extended emission, covering 540\,deg$^2$ and $1.9<z<3.5$ with surface-brightness sensitivities of $\sim2$–$5\times10^{-18}$\,\sbunitsraw. HETDEX aims to constrain the Hubble parameter $H(z)$ and angular diameter distance $D_A(z)$ to better than 1\% precision at $z\sim2.4$, using LAEs as biased tracers of the underlying dark matter distribution \citep{Shoji2009}. To achieve this goal, the project surveys 540\,deg$^2$—corresponding to a comoving volume of 10.9\,Gpc$^3$—in a non-contiguous tiling optimized for sampling large-scale structure \citep{Chiang2013}. Beyond its cosmological goals, HETDEX’s sensitivity and area enable a unique statistical investigation of extended \lya\ nebulae. Despite the low signal-to-noise ratio of most detected LAEs, a subset exhibit spatially resolved emission, providing an unprecedented opportunity to study the frequency, morphology, and environmental dependence of extended \lya\ structures while minimizing cosmic variance. This combination of depth, volume, and unbiased selection makes HETDEX uniquely suited to build the first statistical sample of extended \lya\ sources at Cosmic Noon.

In this work, we perform 2D surface-brightness modeling of every LAE in the internal HETDEX Data Release 5 (HDR5) catalog detected at a signal-to-noise ratio greater than six, bridging the populations of compact Ly$\alpha$ halos and large Ly$\alpha$ blobs.  A description of a smaller subset of this catalog is described in \citet{emc2023} and the full catalog will be provided in an upcoming HETDEX Data Release paper in \citet{hetdex_pdr1}. Section\,\ref{sec:sample} describes the HETDEX observations and sample selection, Section\,\ref{sec:nbimages} outlines the generation of Ly$\alpha$ line-flux maps, and Section\,\ref{sec:fitting} details the surface-brightness model fitting. The accompanying catalog is described in Section\,\ref{sec:cat} and includes positions, redshifts, and morphological parameters for both compact and extended systems, providing the largest homogeneous dataset of its kind. A summary of properties and statistical insights gained from the sample is provided in Section\,\ref{sec:results}.

All positions reported in this paper are in the International Celestial Reference System (ICRS)\null.  We adopt the flat $\Lambda$-cold-dark-matter cosmology with $H_0=67.7\,\mathrm{km}\,\mathrm{s}^{-1}\,\mathrm{Mpc}^{-1}$ and $\Omega_{\mathrm{m},0}=0.31$ measured by \citet[Planck18]{Planck2018}.  All quoted sizes are expressed as physical transverse distances. All magnitudes are expressed in the AB system \citep{oke1983}. We assume a rest-frame vacuum wavelength of $\lambda=1215.67$~\AA\ for Ly$\alpha$. Observed wavelengths expressed in this paper and associated data products are as measured in air. All redshifts are appropriately calculated for any differences between air and vacuum wavelengths using the standard in \cite{Morton1991}.

\section{Observations and Sample Selection}
\label{sec:sample}

\begin{figure*}[ht]
\centering
    \includegraphics[width=0.77\linewidth, trim = 0.8cm 0.3cm 0cm 0.1cm, clip]{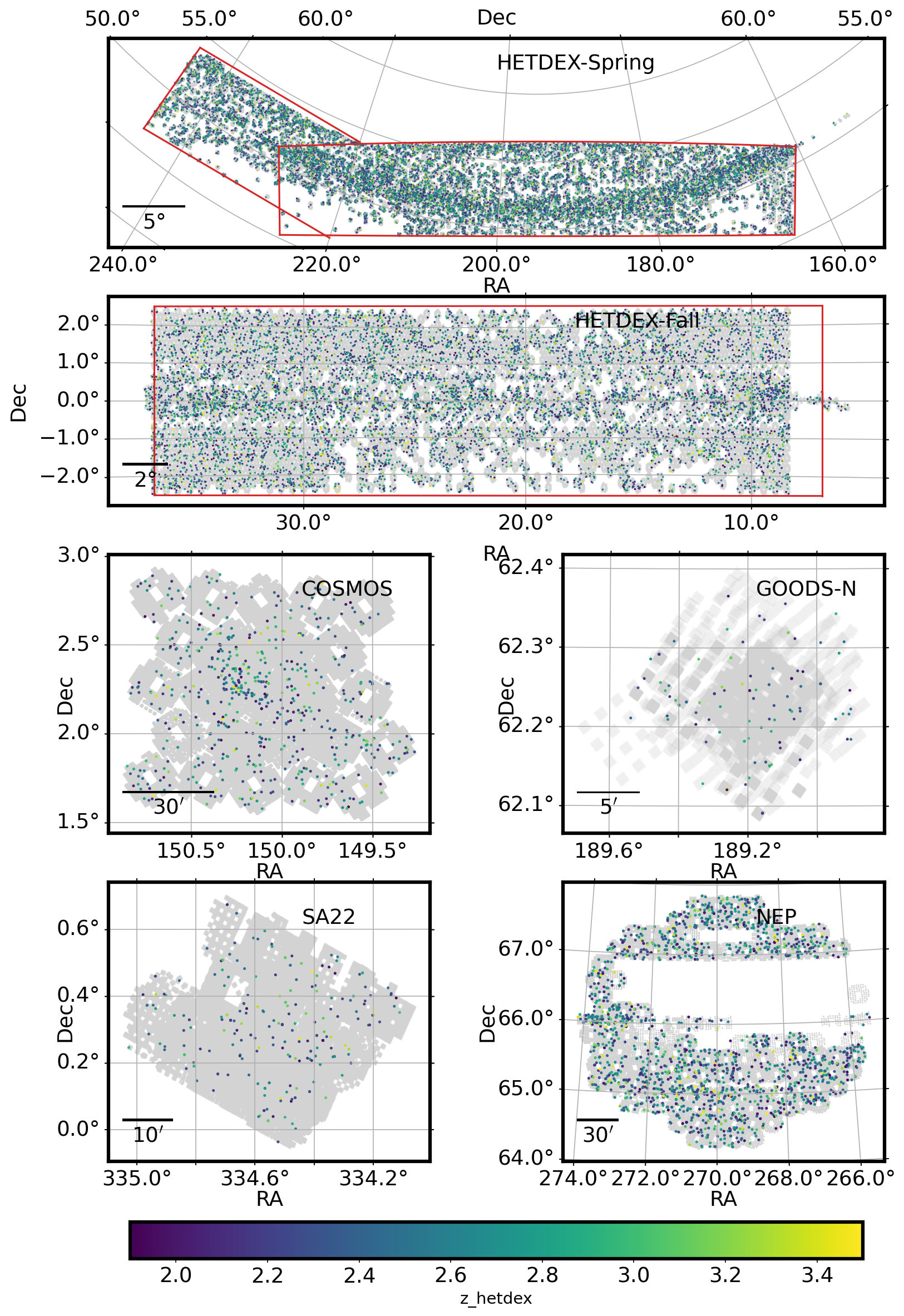}
    \caption{Sky distribution of HETDEX IFU observations and \lya Nebula (LAN) sources across six HETDEX fields. The average sky number density of resolved LAEs is \labdensity\ deg$^{-2}$ (for the redshift range $1.9<z<3.5$ covered by HETDEX). Each panel shows the HETDEX IFU footprints in light gray (with each tile corresponding to a $51\arcsec \times 51\arcsec$ IFU), overlaid with spectroscopically confirmed LANs color coded by redshift (from $\zhet=1.9$ to $\zhet=3.5$). The top two panels display the wide-area Spring (\texttt{dex-spring}) and Fall fields (\texttt{dex-fall}), with rectangular red outlines marking the approximate survey boundaries; these two regions encompass approximately 390 and 150 deg$^2$ of sky area, respectively. The bottom four panels show targeted deep fields: COSMOS, GOODS-N, SA22, and the North Ecliptic Pole (NEP). The scale bars in the lower left of each panel indicate angular size.}
    \label{fig:coverage}
\end{figure*}

\begin{table}[t]
\centering
\small
\begin{tabular}{l@{\hskip 3pt}r@{\hskip 5pt}r@{\hskip 5pt}r@{\hskip 5pt}r@{\hskip 5pt}r}
\hline
Field & $N(\mathrm{IFU})$ & Area & $N_{\mathrm{LAE}}$ & $N_{\mathrm{LAN}}$ & $\Sigma_{\mathrm{LAN}}$ \\
      &                   & (deg$^2$) & & & (deg$^{-2}$) \\
\hline
\hline
DEX-SPRING & 222,822 & 44.72  & 50,195 & 21,563 & 482.19 \\
DEX-FALL   & 126,454 & 25.38  & 20,221 &   9206 & 362.75 \\
NEP        & 31,707  &  6.36  &   4708 &   1851 & 290.89 \\
COSMOS     & 10,139  &  2.03  &   1702 &    648 & 318.44 \\
SSA22      &   4142  &  0.83  &    531 &    230 & 276.69 \\
GOODS-N    &    648  &  0.13  &    265 &    114 & 876.56 \\
\hline
\textbf{TOTAL} & \textbf{395,911} & \textbf{79.46} & \textbf{77,622} & \textbf{33,612} & \textbf{423.02} \\
\hline
\end{tabular}
\caption{Summary of HETDEX field coverage sorted by area. Columns include the number of IFUs, survey area, number of LAEs and LANs, and the surface density of LANs $\Sigma_{\mathrm{LAN}}$ in units of deg$^{-2}$.}
\label{tab:summary}
\end{table}

HETDEX uses the Visible Integral Field Unit (IFU) Replicable Unit Spectrograph  \citep[VIRUS;][]{Hill2021} on the Hobby Eberly Telescope \citep[HET;][]{Ramsey1998, Hill2021} to search for \lya-emitting galaxies (LAEs) at a redshift of $1.9<z<3.5$. Equipped with an array of 78 IFUs,  VIRUS simultaneously obtains $\sim35$\,K fiber spectra in the wavelength range $3500~\mathrm{\AA} \lesssim \lambda \lesssim 5500~\mathrm{\AA}$ with spectral resolving power $750 \lesssim R \lesssim 950$. HETDEX survey tiling takes three-dithered $\sim6$\,minute exposures to fill in the gaps between fibers on each individual IFU\null. In just 20 minutes, HETDEX/VIRUS can detect over 150 LAEs in a $\sim 55$~arcmin$^2$ area of sky.

Each VIRUS IFU covers a $51\arcsec \times 51\arcsec$ field-of-view and consists of 448 $1\farcs 5$-diameter fibers coupled to a dual-channel spectrograph. Observations are typically executed in a three-point dither pattern with 360 s exposures at each position.  This yields near-complete spatial coverage within the IFU footprints. For further details about the HETDEX survey design and its data products, see \citet{Gebhardt2021}, and for instrumentation details, see \citet{Hill2021}.

The sample used in this study is drawn from HETDEX data collected between 2017 January 1 and 2024 July 31. In total, the dataset includes 6771 individual observations, each composed of a varying number of functional IFUs.  The full sky distribution of IFU observations and the outline of the main field regions are found in Figure\,\ref{fig:coverage}. The HETDEX survey consists of two main fields: the 390~deg$^2$ HETDEX Spring field (``dex-spring'') and the 150~deg$^2$ HETDEX Fall field (``dex-fall''). It also contains non-uniform tiling in several legacy fields.  The most extensive coverage comes from collaborative observations with the Texas Euclid Survey for \lya\ \citep{tesla2023} of the North Ecliptic Pole (NEP). Nearly full field coverage of the central 1~deg$^2$ of the Cosmic Evolution Survey (COSMOS; \citealt{cosmos2007}) is included. As well as sparse coverage of the SA22 \citep{sa221998} and GOODS-N \citep{GOODSN} fields. Field coordinates, coverage area, and the number of IFU observations contained in each field are summarized in Table~\ref{tab:summary}.

Early in the survey, as few as 16 IFUs were installed, with some only partially functional. However, by 2022 the full complement of 78 IFUs was deployed, with an average of 72 IFUs providing science-quality coverage each night. As described in \citet{emc2023}, some data are lost due to detector issues, saturation from bright stars or galaxies, and contamination from meteors and satellite streaks. After quality cuts, a total of 395{,}911 science-grade IFU observations remain. A breakdown by survey field is provided in Table~\ref{tab:summary}.

HETDEX attempts to achieve uniform sensitivity across the survey area by adjusting exposure times based on conditions such as image quality, atmospheric transparency, and instrument throughput. Despite these efforts, the detection limit for \lya emission varies significantly across the dataset, with \lya\ surface-brightness sensitivities ranging from $\sim2-5\times$\sbunits\ for the typical spatial (FWHM $=$ 1\farcs2 to 3\farcs0) and spectral resolution ($\Delta \lambda~=~5.6$\AA) of HETDEX. Fiber-to-fiber and IFU-to-IFU sensitivity variations are also present due to instrumental differences and aging effects across the array. 

\begin{figure*}[t]
    \centering
    \includegraphics[width=0.9\textwidth]{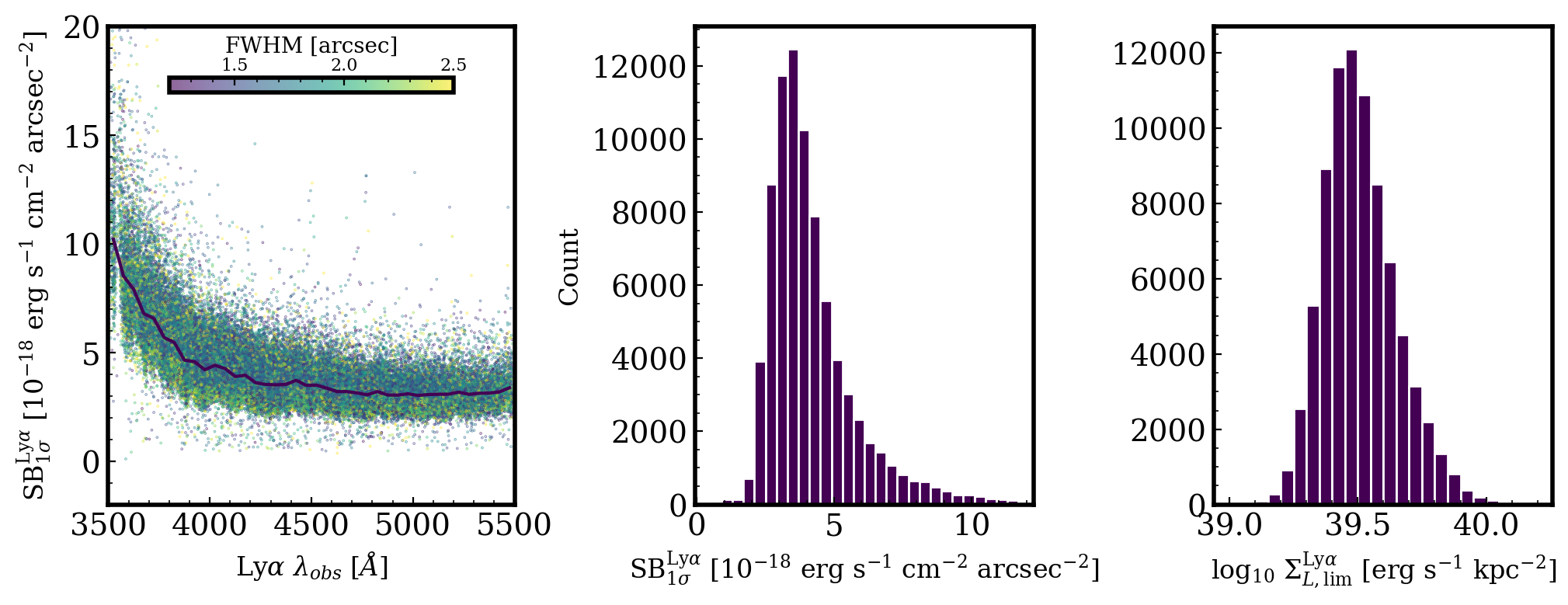}
\caption{$1\sigma$ surface–brightness sensitivity measured from the pixel variance in the continuum-subtracted line-flux maps and converted into intrinsic luminosity surface–density limits for individual HETDEX IFU observations. 
Left: the observed--frame $1\sigma$ Ly$\alpha$ surface brightness, 
$\mathrm{SB}_{1\sigma}^{\mathrm{Ly}\alpha}$ (in units of $10^{-18}$\,erg\,s$^{-1}$\,cm$^{-2}$\,arcsec$^{-2}$), 
as a function of observed wavelength. Each point represents an IFU exposure and is color coded by the image quality (FWHM) in arcseconds. 
The solid line traces the median trend in 50\,\AA\ bins, showing increased sensitivity toward the blue due to higher instrumental throughput and lower sky background. 
Middle: distribution of $\mathrm{SB}_{1\sigma}^{\mathrm{Ly}\alpha}$ values across all exposures, 
typically ranging from 2–10$\times10^{-18}$\,erg\,s$^{-1}$\,cm$^{-2}$\,arcsec$^{-2}$. 
Right: histogram of the intrinsic luminosity surface–density limit, 
$\Sigma_{L,\mathrm{lim}}^{\mathrm{Ly}\alpha}$, derived from each exposure using the luminosity distance and physical scale at its redshift.
These limits correspond to $\log_{10}\Sigma_{L,\mathrm{lim}}^{\mathrm{Ly}\alpha}\sim39.5$–40.1\,erg\,s$^{-1}$\,kpc$^{-2}$, 
illustrating the range of Ly$\alpha$ sensitivity achieved across the survey.
}
    \label{fig:sblim_fwhm}
\end{figure*}

Our analysis sample comes from the fifth internal HETDEX data release (HDR5; v5.0.2 internally), which includes more than 600 million fiber spectra and includes $\sim$78 deg$^2$ of sky.  The full LAE catalog includes over one million candidates and is being prepared for a dedicated HETDEX release paper \citep{hetdex_pdr1}. For selection methods and classification details, see \citet{davis2023} and \citet{emc2023}. From this parent catalog, we select a sample of moderate to high signal-to-noise ratio (S/N $>$ 6) LAEs; objects fainter than this do not have enough counts for surface photometry measurements.  This S/N measure comes from the HETDEX detection pipeline and comes from fitting a Gaussian model to the 1D extracted spectrum. See \citet{Gebhardt2021} for more details on the detection search and emission-line fitting methods.

To ensure robust surface-brightness profile fits, we exclude LAEs that fall within $1\farcs 5$ of an IFU edge, i.e., those with $|x_{\rm ifu}| > 24\farcs0$ or $|y_{\rm ifu}| > 24\farcs0$, where $x_{\rm ifu}$/$y_{\rm ifu}$ is the detection positional distance from the IFU center in arcseconds. Additionally, to minimize contamination from nearby bright-continuum sources, we remove any LAE candidate within 5\arcsec\ of a continuum source with \hetg\ $<$ 20, where \hetg\ is the HETDEX $g$-band magnitude measured by convolving the point-spread function (PSF)-weighted spectrum with the Sload Digital Sky Survey (SDSS) $g$-band filter response. These criteria reduce our final sample to \nlaefull\ LAEs with clean, well-centered line emission.

\section{Line-Flux Maps}\label{sec:nbimages}

\begin{figure*}[t]
    \centering
    \includegraphics[width=0.8\textwidth, trim=0.2cm 0 0 0.3cm, clip]{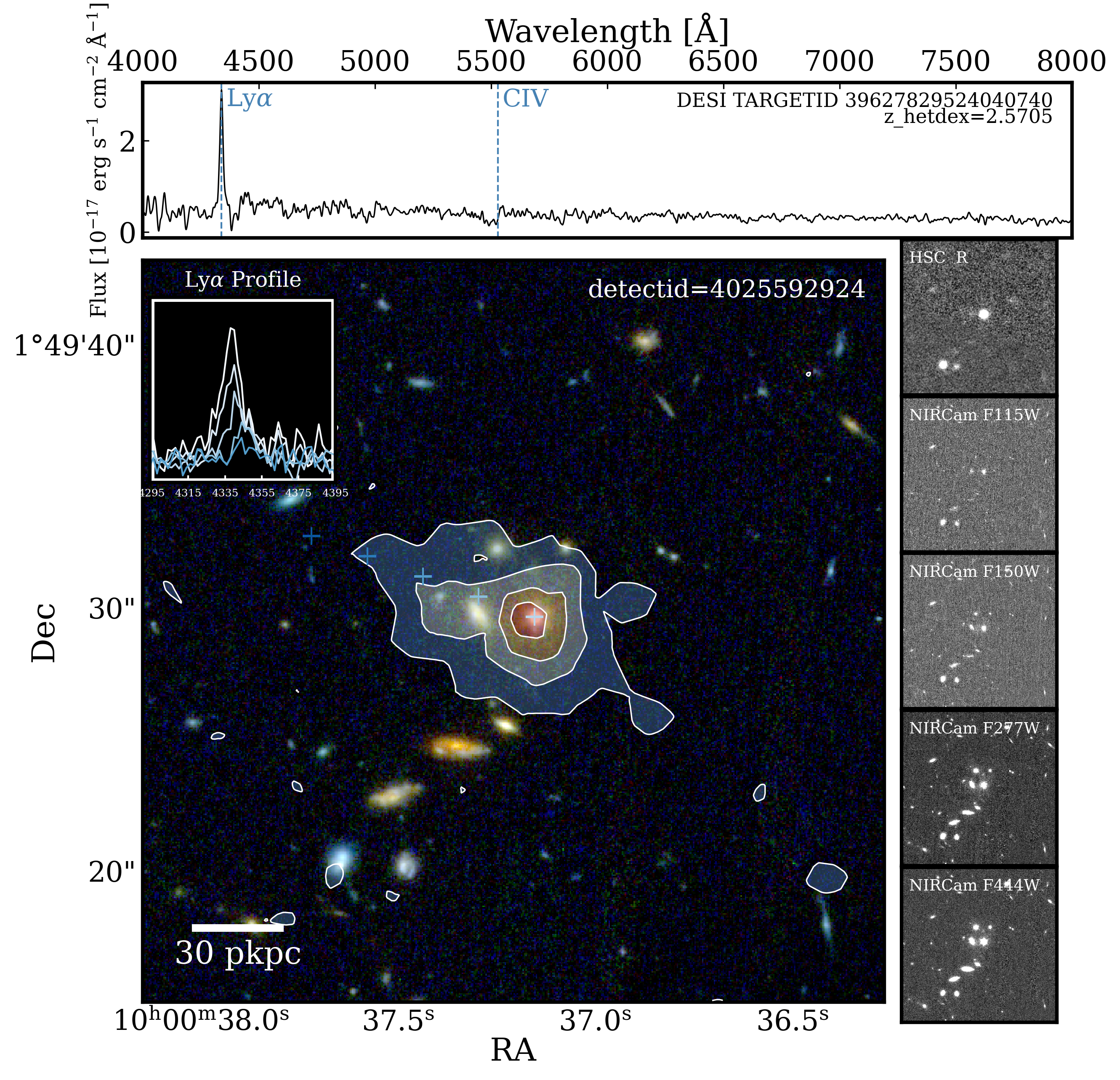}
\caption{Multi-wavelength view of HLAN\,4025592924 at \mbox{$\zhet=2.57$}. Situated in the COSMOS Deep Field, this LAN is among the largest in the HETDEX sample, with an isophotal radius of 45.9\,kpc -- placing it in the top 2\% of the distribution. Despite its size, the structure's \lya\ luminosity is moderate at \texttt{log\_L\_lya} = 43.8\,erg\,s$^{-1}$, and it is not identified as an AGN\null. \textit{Top:}  DESI DR1 (\texttt{TARGETID~39627829524040740}; \citealp{desi2,desidr1}) 1D spectrum on the central bright source (white '+' symbol).  The lack of \ion{C}{4} suggests the source is not AGN dominated.  \textit{Main:}  $30\arcsec\times30\arcsec$ JWST/NIRCam three-colour composite (blue\,=\,$\tfrac12$(F115W\,+\,F150W), green\,=\,F277W,
red\,=\,F444W) from the COSMOS-Web DR1 mosaics \citep{cosmos-webNIRCAM2025}.  The Ly$\alpha$ line-flux map (see Section\,\ref{sec:nbimages}) from the HETDEX data cube is over-plotted as contours (levels $3$–$15\,\sigma$).  
The top-left inset shows 1D Ly$\alpha$ profiles extracted along the positions marked by “+’’ symbols in the image.  A $30$\,kpc scale bar is indicated at lower left. \textit{Right column:} $30\arcsec\times30\arcsec$ postage stamp images (top to bottom) from Subaru/HSC-$r$\,band, and JWST
filters F115W, F150W, F277W and F444W reveal that multiple low-mass galaxies accompany the LAN.}
\label{fig:het_cosmos_rgb}
\end{figure*}

\begin{figure*}[ht]
\centering
\includegraphics[width=0.9\textwidth]{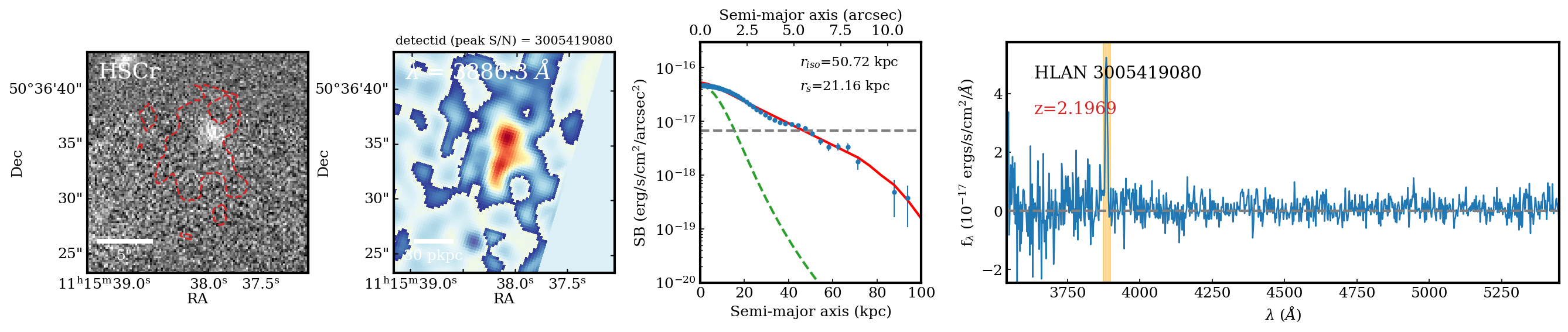}
\includegraphics[width=0.9\textwidth]{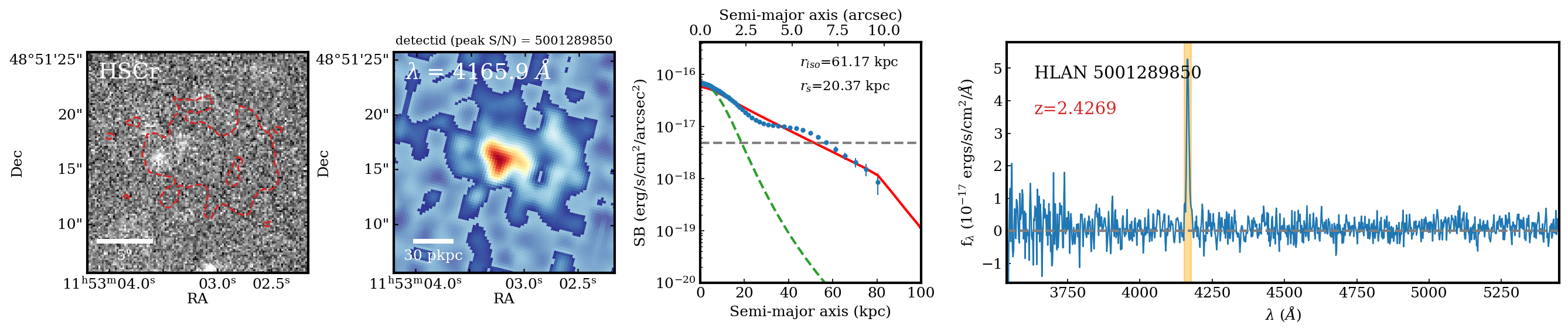}
\includegraphics[width=0.9\textwidth]{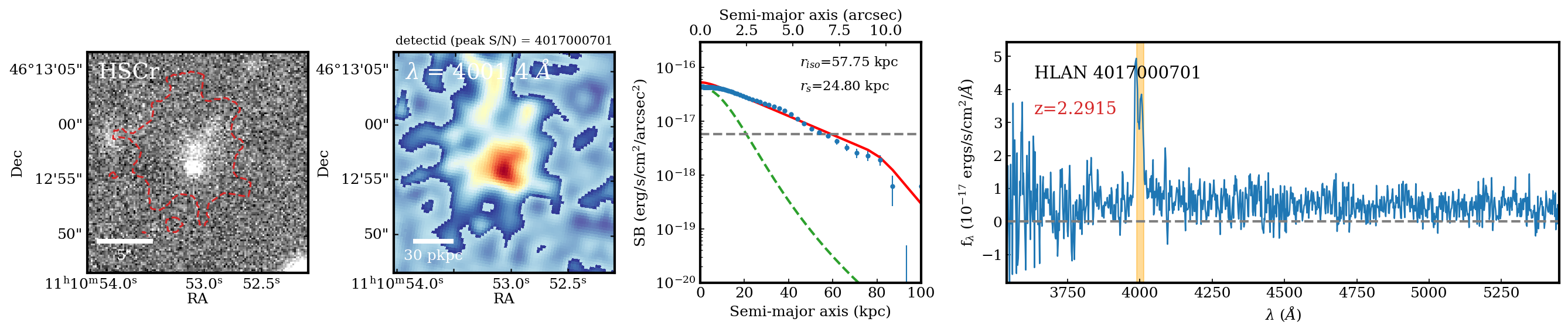}
\includegraphics[width=0.9\textwidth]{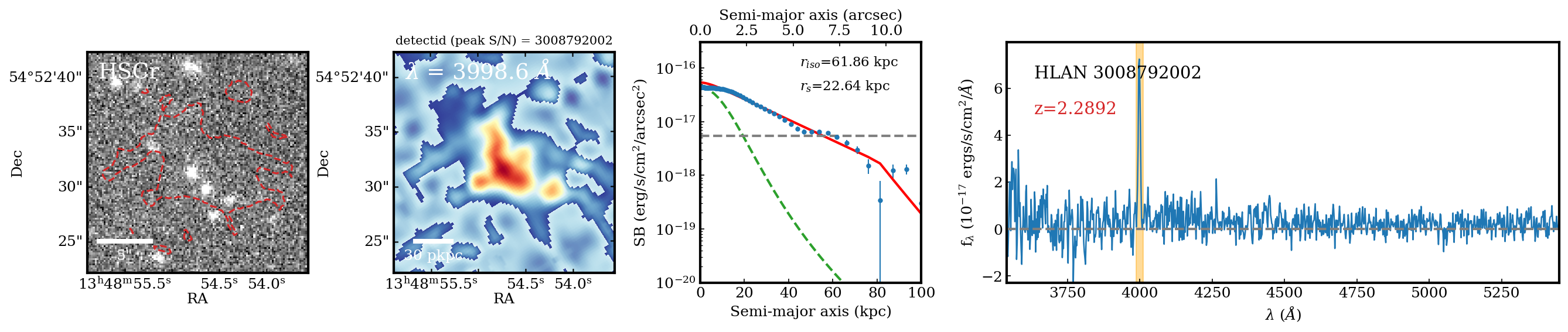}
\includegraphics[width=0.9\textwidth]{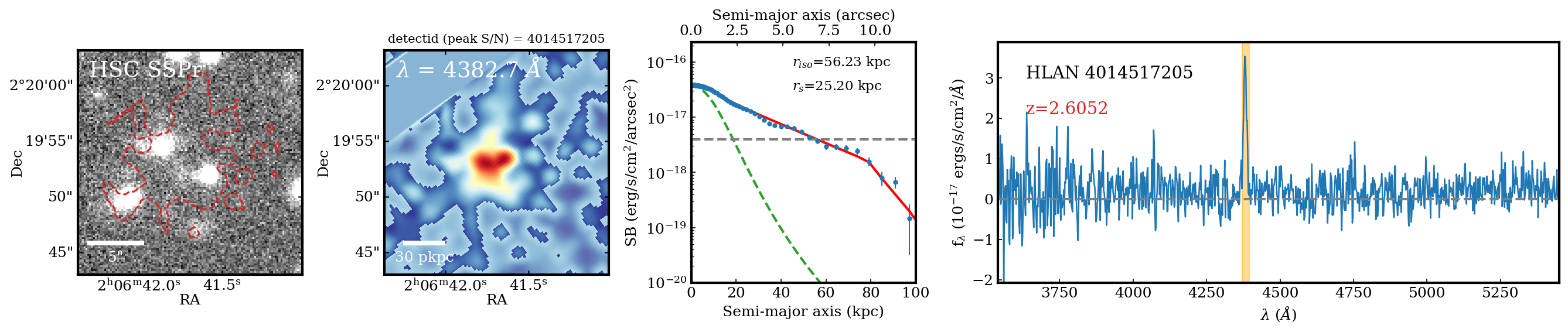}

\caption{Examples of some of the largest extended \lya-emitters. Photometric imaging from HSC-$r$ is shown on the left with the 2$\sigma$ boundary of the \lya emission shown as a dashed red contour. The second column displays the \lya line-flux map centered on the wavelength listed in white text. The third column presents the radial surface-brightness profile in blue and our  best-fit two-component model (PSF core + 2D exponential) in red. The dashed green line is the measured PSF from stars in the same observation as the LAN\null. The HETDEX/VIRUS spectrum for the central HETDEX detection is given in the rightmost panel. This spectrum is the PSF-weighted spectrum from the HETDEX pipeline. The spectral width of the line-flux map is highlighted in yellow on the spectrum. The line shapes are asymmetrical, and some appear to have multiple associated continuum counterparts in HSC-$r$ images.}
\label{fig:ex1}
\end{figure*}

\begin{figure*}[t]
\centering
\includegraphics[width=0.9\textwidth]{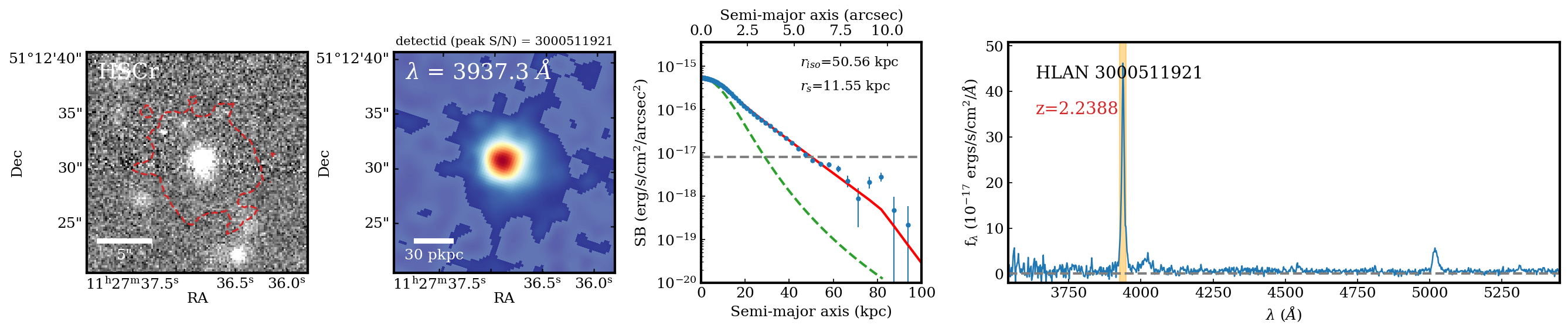}
\includegraphics[width=0.9\textwidth]{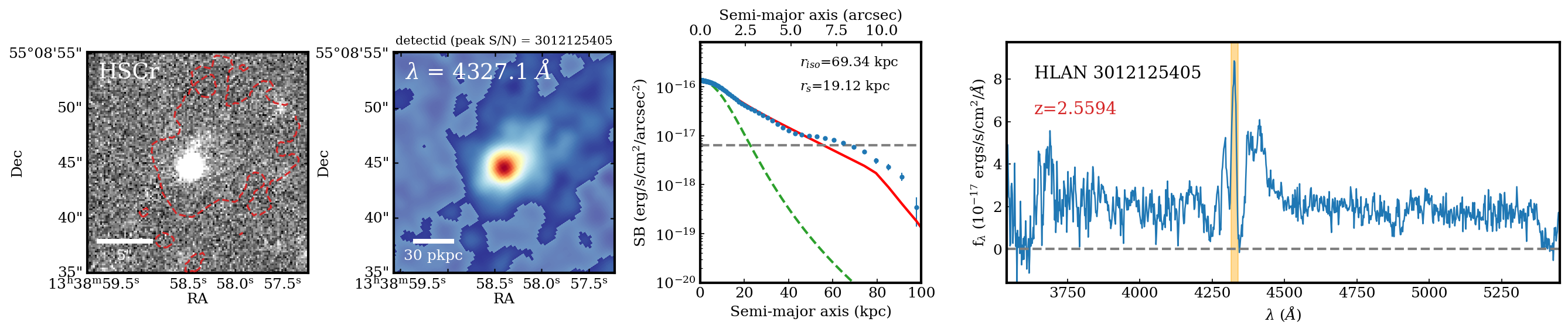}
\includegraphics[width=0.9\textwidth]{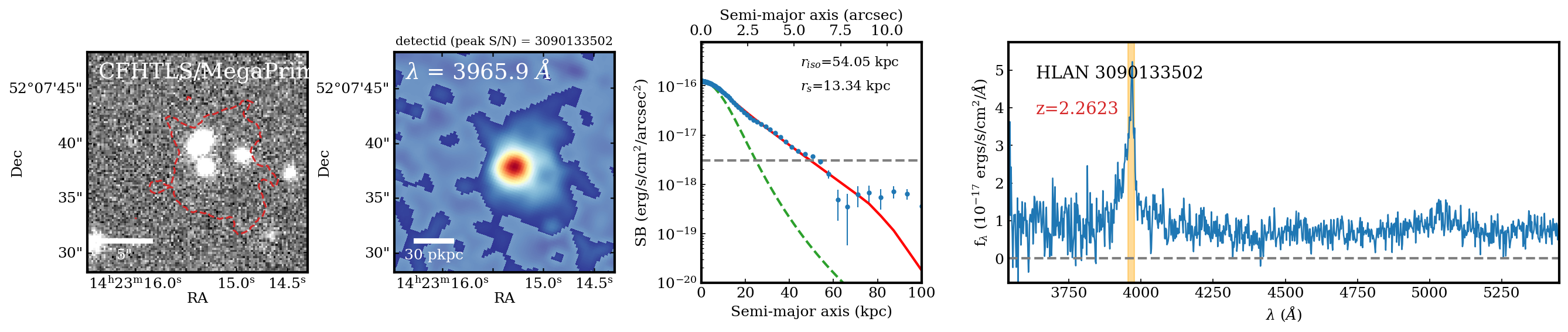}
\includegraphics[width=0.9\textwidth]{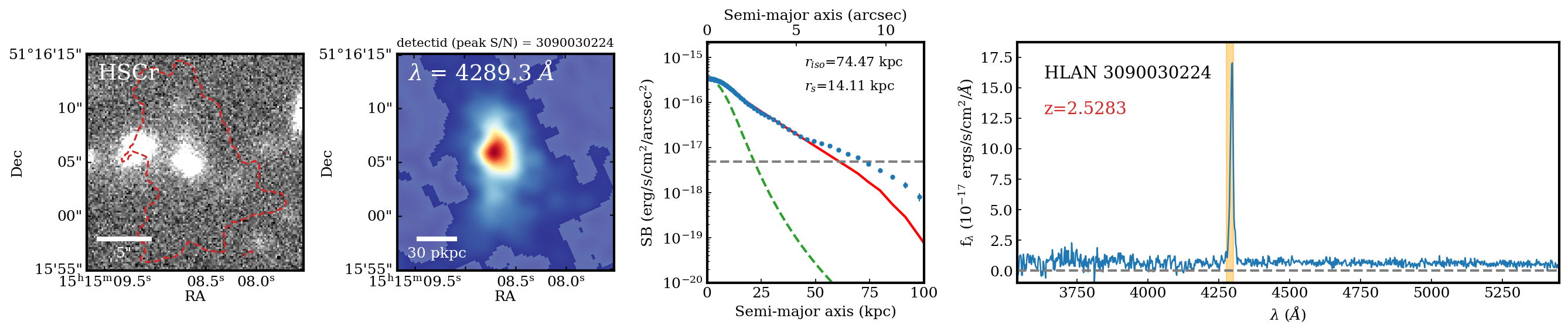}
% These are not 2 component preferred
\includegraphics[width=0.9\textwidth]{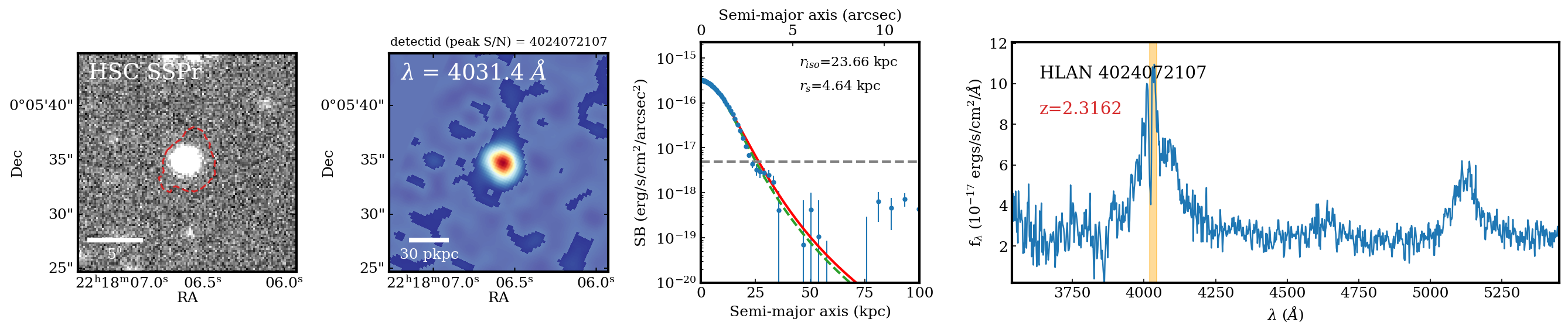}
\includegraphics[width=0.9\textwidth]{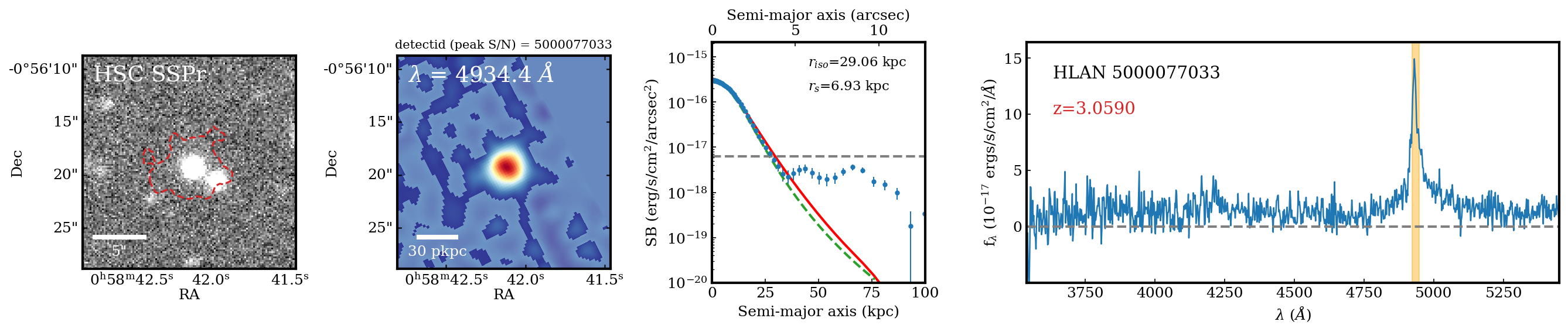}
\caption{Top four panels: examples of extended \lya-emitters with a strong central AGN counterpart. The panels have the same format as described in Figure~\ref{fig:ex1}. Unlike the sources in Figure~\ref{fig:ex1}, the peak \lya-emission coincides with peak continuum emission in accompanying photometric data. Bottom Two Panels: examples of bright AGN that prefer a single-component, point-source model. These are \textbf{not} considered extended by our method.}
\label{fig:ex2}
\end{figure*}

\begin{figure*}[t]

\includegraphics[width=0.9\textwidth]{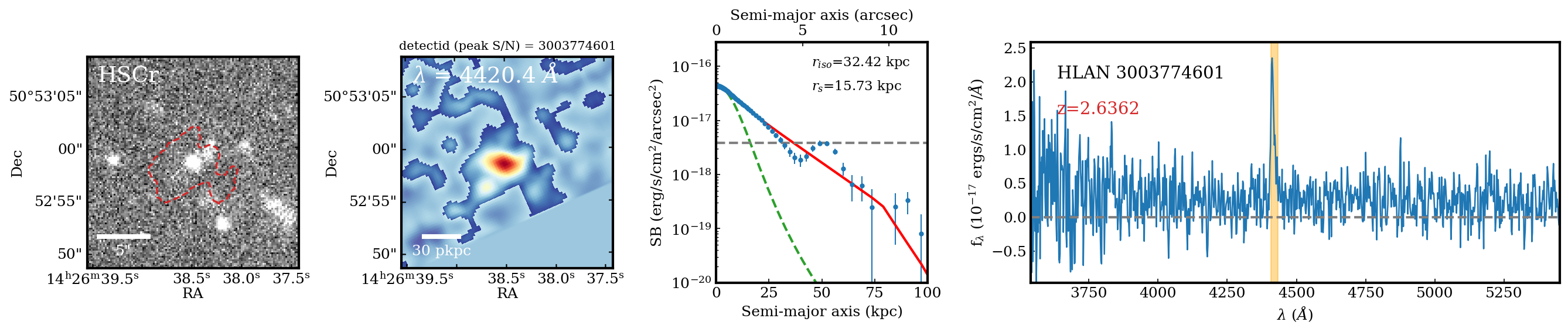}
\includegraphics[width=0.9\textwidth]{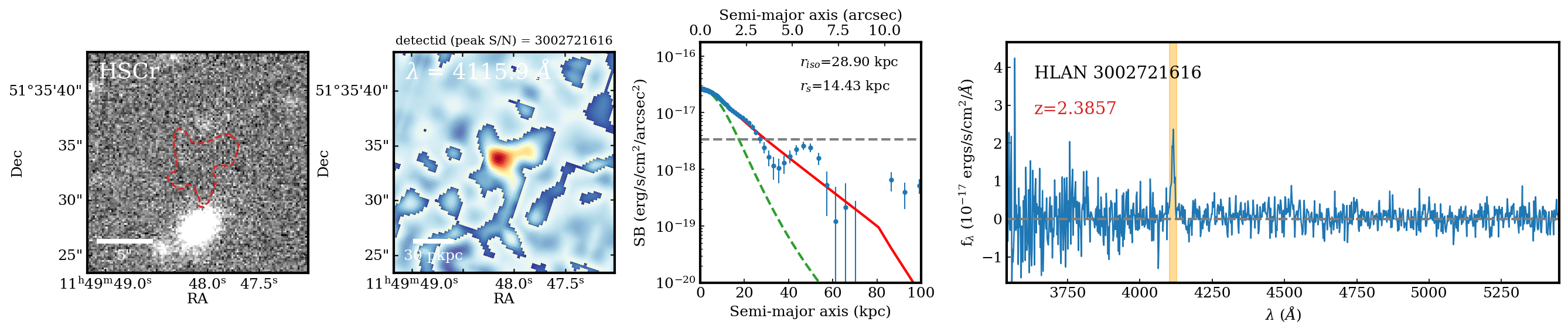}
\includegraphics[width=0.9\textwidth]{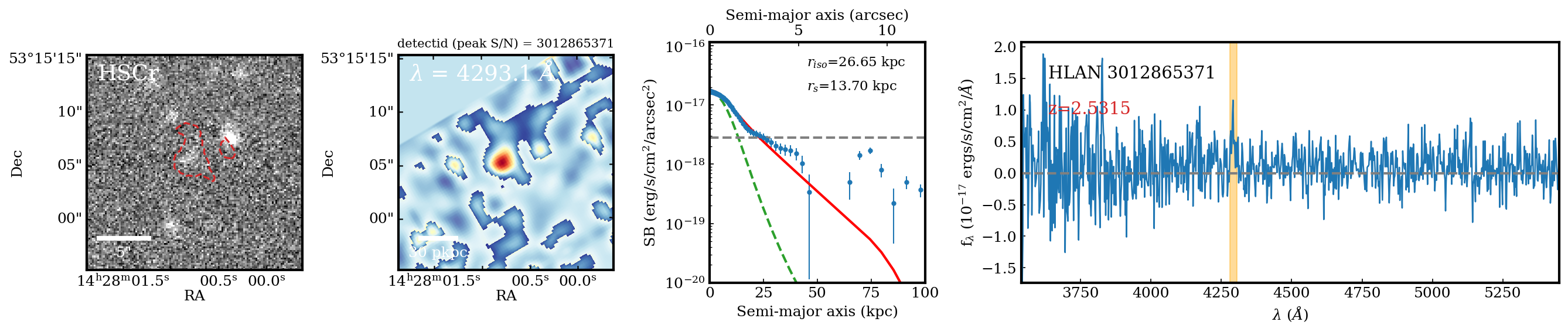}
\includegraphics[width=0.9\textwidth]{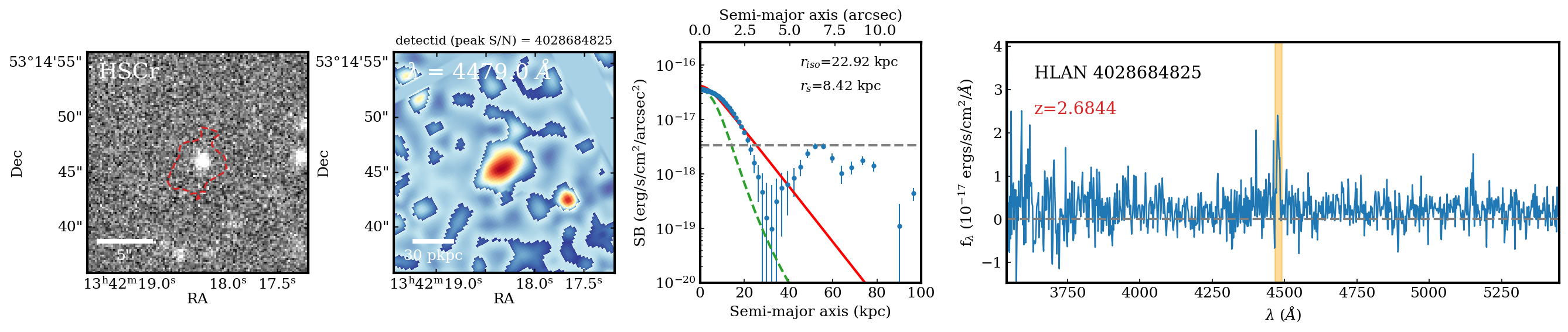}
\includegraphics[width=0.9\textwidth]{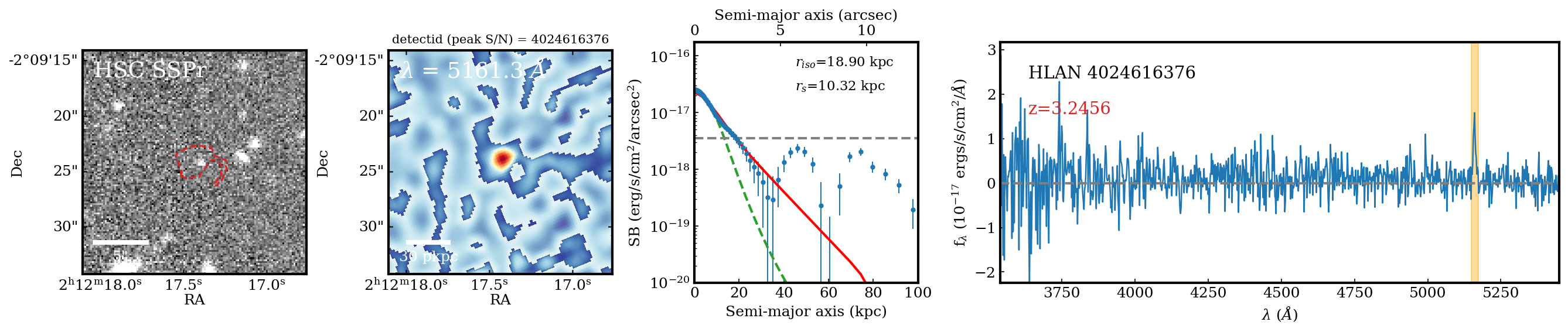}
\includegraphics[width=0.9\textwidth]{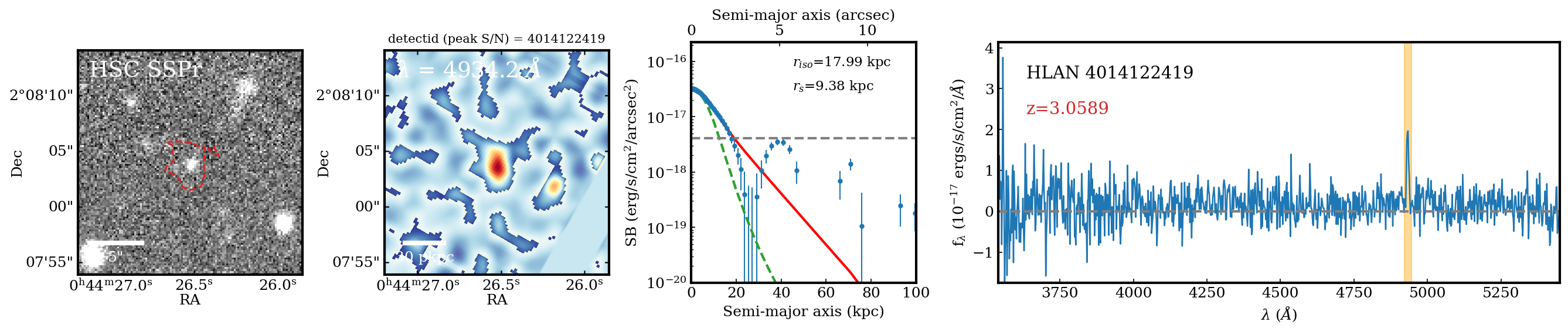}
\caption{Representative examples of smaller LAN in the HETDEX LAE sample. These are plotted from top to bottom in decreasing $r_{\rm iso}$. The panels have the same format as described in Figure~\ref{fig:ex1}.}
\label{fig:ex3}
\end{figure*}

After defining our sample of bright LAEs, we generate their continuum-subtracted \lya line-flux maps. These maps are created using a custom software package, \texttt{hetdex-api}\footnote{https://github.com/HETDEX/hetdex\_api}, developed to work with HETDEX data products.  First, we construct a pseudo-narrowband image at the detected central wavelength of an observed LAE by collapsing the fiber spectra in the spectral dimension around the emission line. This is done by collecting all fibers within a $20\arcsec \times 20\arcsec$ region surrounding the emission-line detection position and taking the summed fiber spectra $\pm2\sigma$ in the wavelength dimension centered on the emission line, where $\sigma$ is the best-fit Gaussian line width derived from the 1D spectral line fit from the HETDEX detection software as described in \citet{Gebhardt2021}. 

Our software flags any fiber spectral elements that are compromised by cosmic rays or other artifacts to NaN. The HET/VIRUS instrument does not contain an atmospheric diffraction corrector, so this correction must be applied in software. While each fiber is assigned a single sky coordinate in the HETDEX data model, the precise (R.A., Decl.) for each element in the fiber spectra actually varies as a function of wavelength relative to its fiducial value of 4500\,\AA\null. Thus a slight correction to the astrometry of up to $\sim$1\arcsec\ across the 3470~\AA\ - 5540~\AA\ range is applied depending on the object's wavelength. We then interpolate the fibers to an image of 0\farcs25 per pixel using the \texttt{cubic} interpolation method from  \texttt{scipy/interpolate/griddata} \citep{scipy} and convert the surface-brightness measurements to units of \sbunits. We repeat this process for the fiber spectra uncertainty arrays to properly track measurement noise. As we collapse the spectra along the wavelength direction, we propagate the uncertainties by summing them in quadrature, ensuring that the final error maps reflect the combined contribution from all spectral channels.

Next, we subtract the continuum from each of these object spectra. The HETDEX data processing pipeline provides two sky-subtraction methods: one with a  ``local'' sky-subtraction model, which uses sky fibers on a single IFU to model the sky, and the other using a ``full-frame`' sky-subtraction model, which considers all IFUs in the observation. (The 78 IFUs, which compose a single HETDEX/VIRUS observation, sample $\sim 16\arcmin \times 16\arcmin$ of sky.)  For our analysis, we use flux-calibrated fibers from the ``local'' sky subtraction model, but because we perform continuum subtraction the choice does not significantly affect our results. Using the local sky, we subtract the spectral continuum in the line-flux maps by generating two additional 50\,\AA-wide wavelength-collapsed images,  $\pm10$\,\AA\ from the emission-line map limits.  We take an average of these two continuum images and subtract it from the \lya\ line-flux map to create a continuum-subtracted emission-line-flux map. For broad-line AGN, we note that this continuum subtraction approach does not explicitly model spectral continuum slopes, but the continuum is estimated independently in each spatial element.

The continuum-subtracted surface-brightness sensitivity, $\mathrm{SB}_{1\sigma}^{\mathrm{Ly}\alpha}$, blueis measured from the variance in the pixels in the continuum-subtracted line-flux map generated for each LAE detection. These maps have a pixel resolution of 0\farcs25 and are collapsed spectrally based on the LAE detection's fitted line width ($\sigma=1.8-50$\AA, $\sigma_{\rm avg}=4.4$\AA). The value varies depending on observing conditions and wavelength, as well as subtle variations in the detector response. Figure~\ref{fig:sblim_fwhm} (left) shows $\mathrm{SB}_{1\sigma}^{\mathrm{Ly}\alpha}$ as a function of observed \lya\ wavelength, color coded by the seeing full width at half-maximum (FWHM) for each exposure. The median trend (solid line) reveals a gradual increase in background noise toward shorter wavelengths. The median sensitivity across 
the survey is $3.8\times10^{-18}$\,erg\,s$^{-1}$\,cm$^{-2}$\,arcsec$^{-2}$, with 50\% range of $3.1$–$4.8\times10^{-18}$\,erg\,s$^{-1}$\,cm$^{-2}$\,arcsec$^{-2}$.
The sensitivity improves significantly at longer observed wavelengths where the instrument throughput is higher, and the sky background is lower. For wavelengths $\lambda_{\mathrm{obs}}>4000$\,\AA, corresponding to 
$z_{\mathrm{Ly}\alpha}>2.29$, the median surface--brightness limit is $3.4\times10^{-18}$\,erg\,s$^{-1}$\,cm$^{-2}$\,arcsec$^{-2}$, 
whereas for $\lambda_{\mathrm{obs}}<4000$\,\AA\ ($z_{\mathrm{Ly}\alpha}<2.29$) 
it degrades to $5.3\times10^{-18}$\,erg\,s$^{-1}$\,cm$^{-2}$\,arcsec$^{-2}$ 
due to reduced throughput and higher detector noise.  
  
The corresponding intrinsic luminosity surface--density limit, $\Sigma_{L,\mathrm{lim}}^{\mathrm{Ly}\alpha}$, is derived for each exposure using the 
luminosity distance and angular--diameter scale at its redshift. The distribution, shown in the rightmost panel of Figure~\ref{fig:sblim_fwhm}, of $\log_{10}\Sigma_{L,\mathrm{lim}}^{\mathrm{Ly}\alpha}$ peaks near 39.5\,erg\,s$^{-1}$\,kpc$^{-2}$ with an interquartile range of 39.4–39.6\,erg\,s$^{-1}$\,kpc$^{-2}$.

An example line-flux map from \texttt{HLAN\,4025592924} is displayed in Figure~\ref{fig:het_cosmos_rgb} as a contour overlay. The source lies at \zhet=2.57 with \lya-emission at
$\lambda$=4344\,\AA\null.  This source's size is in the top 2\% of our LANs, with an effective isophotal radius, $r_{\rm iso}$, of 45.9\,kpc. An inset in the main panel displays an expanded view of the line profile at seven apertures across the nebula, revealing the nebula's velocity structure. This object is representative of the larger LANs in the sample, but is special in that it lies in the COSMOS-Web field \citep{cosmoswebmain}, has sensitive, high spatial resolution data in the near-infrared from JWST NIRCam \citep{cosmos-webNIRCAM2025}, and high spectral resolution spectrum from DESI DR1 \citep{desi2022, desidr1}. The DESI DR1 coadded spectrum (top row) clearly shows broad Ly$\alpha$ but the absence of the \ion{C}{4} doublet suggests no strong AGN activity. The three-color JWST composite reveals multiple possible counterparts, but only one is bright in the Hyper Suprime-Cam $r$-band \citep[HSC-$r$;][]{hsc}. The Ly$\alpha$ halo extends well beyond the stellar continuum.

Further examples of the line-flux maps are shown in the second column panel in  Figures~\ref{fig:ex1} -- \ref{fig:ex3}. The leftmost panels in both figures, matched in coordinate space, show supplemental $r$-band images from the HSC\null. The dashed red contours denote the LANs' 2$\sigma$ isophotes.

\section{Model Fitting}
\label{sec:fitting}

For every \lya line-flux map in the sample, we determine whether a source's \lya-emission is extended by comparing two surface-brightness models produced by the software package \texttt{pyimfit/imfit} \citep{Erwin2015}. The first model represents point-source emission, as determined by nearby field stars; the second  includes this point source, but also adds in a symmetric exponential component to model the extended emission.

\subsection{Point-source Model}

The impact of the atmosphere and the HET/VIRUS optical path is well described by a 1D Moffat\footnote{} function \citep{Moffat1969} with $\beta=3.5$ \citep{Hill2021}. 
\begin{equation}
I(\theta) = \frac{I_0}{\left(1 + (\theta/\alpha)^2\right)^\beta}
\label{eq:moffat}
\end{equation}
where $\theta$ is the angular separation on the sky, $I_0$ is the central intensity, $\beta$, fixed to 3.5, determines the steepness of the wings, and $\alpha$ defines the width of the profile, via
\begin{equation}
\alpha = \frac{\mathrm{FWHM}}{2\sqrt{2^{1/\beta} - 1}}
\label{eq:alpha_fwhm}
\end{equation}

For each HETDEX observation, the image quality or FWHM is determined as part of the reduction pipeline described by \citet{Gebhardt2021}.  An average value across all stars in a single HETDEX observation (typically 20-30 stars) is used as a representative value for each observation and is stored in each internal HETDEX data release. 

%this work is done in /work2/05350/ecooper/stampede2/psf_fitting/psf_fitting.ipynb
To test for possible differences between the HETDEX pipeline image quality and the image quality applicable to our line-flux maps, we independently assess the latter using spectrally collapsed, interpolated fiber data of stellar spectra. This approach follows the same procedure used to create the line-flux maps described in Section~\ref{sec:nbimages}, except that the spectrum is collapsed over the wavelength range from 4000 to 5000~\AA\null. We compute the mean, median, and standard deviation of the collapsed image and subtract the mean sky to normalize the background. Pixels located beyond a radius of 8\arcsec\ from the star's center and exceeding 5 times the sky background level are masked, as they are likely contaminated by background sources.

Using the software \texttt{pyimfit/imfit}\footnote{\url{https://www.mpe.mpg.de/~erwin/code/imfit/}} \citep{Erwin2015}, we fit a symmetric 2D Moffat profile with fixed $\beta = 3.5$ to each star with a \hetg\ magnitude between $\sim 16$ and 21. The intensity and FWHM of the Moffat function are treated as free parameters. In total, 42,923 stars are analyzed.

The PSF measurements derived from this approach exhibit a systematic offset of +0\farcs14 compared to the FWHM measurements obtained from the HETDEX reduction pipeline \citep{Gebhardt2021}. This offset is smaller than the standard deviation between the two measurements (0\farcs23), and can likely be explained by  minor differences in masking, image interpolation, normalization, and fitting procedures.

To incorporate this offset into our PSF modeling, we adjust the FWHM used in subsequent fits. When modeling both a simple point source and the core component of the two-component model, we fix the FWHM to the value measured in the reduction pipeline, with an additional 0\farcs14 added to account for the systematic offset described above. 

Additionally, the modeling of the core is assumed to be circular, and the intensity is treated as a free parameter. The spatial center of the model is allowed to vary within $\pm5$\arcsec\ of the central coordinate reported by the HETDEX detection. This offset is intentionally large to account for any possible astrometric errors due to detection grouping.

\subsection{Core Plus Exponential Model}

Building on the simple point-source model, we implement a two-component approach that includes both a compact core and an extended exponential halo, a common approach adopted for LAHs (for example, in \citealt{Wisotzki2016}). The core is modeled as a point source using the same method described above. This is a reasonable assumption, as the typical seeing of HETDEX observations ($1\farcs 2$ to $3\farcs 0$, or $\sim 10$ to 25~kpc at $z \sim 2.5)$ is comfortably larger than the UV continuum core-scale lengths for LAEs in the Hubble Ultra-Deep Field, constrained by Hubble Space Telescope imaging \citep{Leclercq2017}.

%While LANs display a variety of morphologies, our ability to constrain source ellipticity is highly dependent on per-pixel signal-to-noise ratios.  A few low S/N pixels far from the source center can force fitting routines to return artificially high ellipticity values. Since asymmetric modeling is only viable for the brightest and most extended emitters, the inclusion of these additional parameters is beyond the scope of this work.  By assuming circularly symmetry, we reduce the dimensionality of our parameter space and allow more robust comparisons with the single-component PSF model. %As with the one-component model, the spatial center of the model is allowed to vary within $\pm5$\arcsec\ of the central coordinate reported by the HETDEX detection.  

While LANs display a variety of morphologies, the signal-to-noise of individual surface-brightness maps is insufficient for morphological analysis for most objects in our sample. Instead, we only focus on the circularly averaged extent through our two-component surface-brightness model defined as

\begin{equation}
\label{eq:2comp}
I(\theta) = I_{\rm core}(\theta) +  I_{\rm exp} \exp\left(-\frac{\theta}{\theta_s}\right)
\end{equation}
where $I_{core}(\theta)$ is defined by Equations~\ref{eq:moffat}~and~\ref{eq:alpha_fwhm} (using the FWHM from the HETDEX reduction pipeline increased by +0\farcs14 to correct for the systematic offset described above) and $\theta_s$ is the scale length of the exponential. After fitting, we report the scale lengths in physical kpc.

In our fitting, the intensity of the exponential component, $I_{\rm exp}$, is allowed to vary, but we require that this extended component contribute at least half of the total intensity at $\theta = 0$. This constraint reduces the likelihood of spurious fits in low S/N regions and ensures that a meaningful fraction of the total light comes from the exponential halo. There is evidence, particularly among sources with strong continuum counterparts, that supports a population of objects with a strong central core component plus a moderately bright exponential envelope that contributes less than half of the total intensity. An example of such a source is found in the bottom panel in Figure~\ref{fig:ex2}. Detailed analysis of the surface-brightness profiles of AGN will follow in future HETDEX papers.

\subsection{Fitting Method and Fit Quality}
\label{sec:fitquality}

We fit every \lya\ line-flux map in the sample with both a core and a core-plus-exponential model using \texttt{pyimfit} \citep{Erwin2015}. To estimate the uncertainties in the fitted parameters, we perform 20 bootstrap iterations for each fit.

Two criteria are applied to evaluate the quality of the fits, including the reduced chi-squared statistic and sufficient signal-to-noise:

\begin{enumerate}
    \item \textit{\lya\ Flux Significance:} The flux measured within the isophotal aperture must be detected at greater than $3\sigma$, i.e.\ \texttt{flux\_lya}~$>3\times$\texttt{flux\_lya\_err}.
    
    \item \textit{Goodness of Fit:} The reduced chi-squared value of the exponential model must satisfy \texttt{chi2\_exp\_reduced}$<3$, or $3\le$~\texttt{chi2\_exp\_reduced}$<5$ for sources with exceptionally high signal-to-noise, defined as \texttt{flux\_lya}/\texttt{flux\_lya\_err}$>300$. 

\end{enumerate}

Applying these quality cuts to the full sample of \nlaefull\ LAEs yields a final sample of \nlaegood\ sources.

\subsection{Determining Size Significance}
\label{sec:resolved}

With the PSF-only model nested within the PSF$+$exponential model, we opt to use a classical nested-model $F$-test \citep{press2007numerical} to assess whether adding the exponential halo yields a statistically significant improvement as opposed to a least-squares likelihood approach. We use this in combination with a difference in Bayesian information criterion ($\Delta$BIC; \citealt{liddle2007}) as well as other significance measures to determine whether a source is extended.

The F-statistic measures how much $\chi^2$ improves per additional parameter, relative to the residual variance of the more complex model.

\begin{equation}
\label{eq:ftest}
F \;=\; 
\frac{\left(\chi^2_{\mathrm{psf}} - \chi^2_{\mathrm{exp}}\right) / (k_{\mathrm{exp}} - k_{\mathrm{psf}})}
     {\chi^2_{\mathrm{exp}} / \nu_2},
\end{equation}

where $k_{\mathrm{psf}}$ and $k_{\mathrm{exp}}$ are the numbers of free parameters in the PSF and PSF$+$Exp models, respectively, and $\nu_i \equiv N_{\mathrm{eff}} - k_i$ are their degrees of freedom, with $N_{\mathrm{eff}}$ denoting the number of unmasked pixels used in the fit. The full chi-square values are denoted $\chi^2_{\mathrm{psf}}$ (PSF) and $\chi^2_{\mathrm{exp}}$ (PSF$+$Exp).

The corresponding $p$-value, $p_F$, quantifies the probability of obtaining an $F$-statistic as large as or larger than the observed value under the null hypothesis that the PSF-only model is sufficient. It is computed from the cumulative distribution function (CDF) of the $F$ distribution as
\begin{equation}
p_F = 1 - {\rm CDF}_F(F;\,k_{\mathrm{exp}}-k_{\mathrm{psf}},\,\nu_{\mathrm{exp}}),
\end{equation}
where $(k_{\mathrm{exp}}-k_{\mathrm{psf}})$ and $\nu_{\mathrm{exp}}$ are the numerator and denominator degrees of freedom, respectively. In practice, a small value of $p_F$ (e.g.\,$p_F<0.05$) indicates that the extended model provides a statistically significant improvement in the fit, whereas a large $p_F$ suggests that the simpler PSF-only model is sufficient.

In addition to the $F$-statistic, we also compute the difference in Bayesian Information Criterion ($\Delta$BIC; \citealt{liddle2007}) between the PSF‐only and PSF+Exp models, defined as 

\begin{equation}
\Delta \text{BIC} = \text{BIC}_{\text{PSF+Exp}} - \text{BIC}_{\text{PSF}},
\end{equation}

where Bayesian Information Criterion (BIC; \citealt{schwarz1978}) is defined as

\begin{equation}
\label{eq:bic}
\mathrm{BIC} = \chi^2 + k \ln(N_{\mathrm{eff}}),
\end{equation}

where $\chi^2$ is the full chi-square value of the fit, $k$ is the number of free parameters in the model, and $N_{\mathrm{eff}}$ denotes the number of unmasked pixels used in the fit. This expression follows from the general definition $\mathrm{BIC} = -2\ln\hat{L} + k\ln N_{\mathrm{eff}}$ under the assumption of Gaussian-distributed residuals with approximately constant variance, for which $-2\ln\hat{L}$ reduces to $\chi^2$ up to an additive constant. The Gaussian likelihood approximation is well justified for our fits, as the pixel-level uncertainties in the model images are dominated by nearly Gaussian detector and sky noise. 

Differences in BIC are widely used as approximate measures of relative model evidence, with values of $|\Delta\mathrm{BIC}|\gtrsim 6$--10 often interpreted as strong to very strong support for the model with lower BIC (e.g., \citealt{kass1995,liddle2007}). In this work, we adopt a slightly more permissive threshold of $\Delta\mathrm{BIC}\le -5$ to identify extended sources. This choice is empirically motivated by visual inspection of all objects with $-10 < \Delta\mathrm{BIC} < 0$, which shows that sources with $\Delta\mathrm{BIC}\le -5$ consistently exhibit clear spatial extension relative to the PSF, while objects with less negative values are increasingly ambiguous.

We find one additional criterion to be useful for determining whether a source is resolved. We compare the measured isophotal radius, $r_\mathrm{iso}$, to that predicted by the best-fit exponential model based on the surface-brightness limit of each image. The predicted isophote is computed as

\begin{equation}
r_{\mathrm{iso,pred}} = \frac{r_\mathrm{e}}{1.678} 
    \ln \left( \frac{I_0}{2\,\sigma_{\mathrm{pix}}} \right),
\end{equation}

where $I_0$ is the model’s central surface brightness and $\sigma_{\mathrm{pix}}$ is the background RMS per pixel, derived from the measured sky noise (\texttt{SB\_1sigma\_obs}) and pixel scale (0\farcs25).  

We then quantify the fractional mismatch between the measured and predicted isophotal radii as

\begin{equation}
\mathrm{iso\_rel\_err} = 
\frac{|r_\mathrm{iso} - r_{\mathrm{iso,pred}}|}{r_{\mathrm{iso,pred}}}.
\end{equation}

This quantity, stored as the catalog column \texttt{iso\_rel\_err}, measures the relative deviation between the observed morphology and that expected from the best-fit exponential profile. Sources with $\mathrm{iso\_rel\_err} > 1$ are flagged as inconsistent (\texttt{flag\_iso\_mismatch}) and excluded from the extended-source sample, as such discrepancies typically indicate unstable or non-physical fits generally due to low signal-to-noise or from contamination of foreground objects. This test provides an additional safeguard against false extensions, ensuring that only objects whose observed isophotal radii are consistent with their modeled surface-brightness profiles are classified as genuinely extended. Values of \texttt{iso\_rel\_err} close to zero indicate good agreement between the observed and model-predicted isophotal radii, while large values signify inconsistency between the fitted profile and the measured morphology.  

A source is deemed to be the best fit by a two–component model when the following quantitative criteria are satisfied:

\begin{enumerate}
    \item F-test criterion: \texttt{log10\_pF}~$<-12$.
    \item $\Delta$BIC criterion: \texttt{dBIC}~$<-5$
    \item Isophotal consistency: The observed and model-predicted isophotal radii must agree within a factor of two (\texttt{iso\_rel\_err}~$<1$).
    \item $r_\mathrm{s}$ significance: To ensure the halo is genuinely resolved, we additionally require that the exponential scale length is significantly detected, \texttt{r\_s}~$>3~\times$~\texttt{r\_s\_err}.
\end{enumerate}

Out of the sample of \nlaegood\ sources that satisfy the quality criteria described in Section\,\ref{sec:fitquality}, \nlan\ are best fit by a two-component, extended model, indicating that nearly half of the sample exhibits statistically significant extended \lya\ emission in which the exponential halo contributes at least half of the total model intensity.

\subsection{Isophotal Effective Radius}

We measure the radial surface–brightness profile of each source using a sequence of circular annuli centered on the best–fit emission peak of the two–component model. These profiles provide a direct, model–independent view of the \lya\ surface–brightness decline. Examples are shown in the third column of Figures~\ref{fig:ex1}--\ref{fig:ex3}, where blue points indicate the radial profile and the horizontal dashed line marks the standard deviation of the sky background, $\sigma_{\mathrm{sky}}$. %Beyond $\theta > 12\arcsec$, we mask any spurious features exceeding five times the background variance.

The isophotal radius, $r_{\mathrm{iso}}$ (catalog column \texttt{r\_iso}), is defined as the radius at which the azimuthally averaged \lya\ surface brightness falls below the local background standard deviation, $\sigma_{\mathrm{sky}}$ (catalog column \texttt{SB\_1sigma\_obs}. The corresponding isophotal flux (\texttt{flux\_lya}) and circularized area (\texttt{area\_r\_iso\_circ}, in arcsec$^2$) are measured within the aperture defined by $r_{\mathrm{iso}}$ and listed in Table~\ref{tab:main}.

This definition differs from the isophotal contour–based approaches used in previous work \citep[e.g.,][]{matsuda11, cantalupo14, hennawi15, cai17}, which are circularized from the contour area and converted to an effective radius. Our method instead derives $r_{\mathrm{iso}}$ directly from the azimuthally averaged profile, ensuring a uniform and reproducible measurement across the full sample. The difference arises from methodology rather than PSF or sensitivity variations: our profiles extend to the $1\sigma$ surface–brightness limit and neglect ellipticity, typically yielding slightly larger radii than contour–based measurements. This approach is particularly robust for faint or low–S/N halos where 2D contouring becomes unreliable and averaging the signal in annuli provides higher S/N. Although our circular–profile method yields systematically larger effective radii, it provides a consistent comparison to surface–brightness analyses of lower–luminosity halos from MUSE surveys \citep[e.g.,][]{Wisotzki2016, Leclercq2017} and stacked HETDEX samples \citep{LujanNiemeyer22a, LujanNiemeyer22, McKay2025}.

\subsection{Isophotal Area}

To compare our approach with standard isocontour methods, we also measure the isophotal area, $A_{\mathrm{iso}}$, directly from the \lya\ line–flux maps. The area enclosed by the 2$\sigma$ surface–brightness contour—corresponding to twice the observed 1$\sigma$ limit (\texttt{SB\_1sigma\_obs})—defines \texttt{area\_iso\_2sigma} in arcsec$^2$. This quantity allows us to compare contour–based areas with those inferred from the circularized isophotal radius. Both metrics are included in the catalog as described in Table~\ref{tab:main}. The comparison is discussed in detail in Section~\ref{sec:area_comparison}.

\section{Catalog Description}
\label{sec:cat}
\begin{table*}[ht]
\centering
\footnotesize
\setlength{\tabcolsep}{4pt} % optional
\begin{tabular}{lll}
\hline
\textbf{Column Name} & \textbf{Unit} & \textbf{Description} \\
\hline
\hline
\texttt{name}               & HLAN$+$\texttt{detectid}               & Unique observation-specific HETDEX \lya\ Nebula (HLAN) identifier. \\
\texttt{ra}                 & deg                                    & Best-fit model coordinate R.A. \\
\texttt{dec}                & deg                                    & Best-fit model decl. \\
\texttt{source\_type}       & \textemdash{}                                      & Source class (\texttt{lae} or \texttt{agn} if in HETDEX AGN catalogs; \citealt{liu2022,liu2025}). \\
\texttt{z\_hetdex}          & \textemdash{}                          & Spectroscopic redshift (see \citealt{emc2023}). \\
\texttt{z\_hetdex\_src}     & \textemdash{}                          & Redshift source as described in \citet{emc2023}. \\
\texttt{detectid}           & \textemdash{}                          & HETDEX unique internal detection ID. \\
\texttt{shotid}             & \textemdash{}                          & HETDEX observation ID. \\
\texttt{field}              & \textemdash{}                          & Survey field label. \\
\texttt{SB\_1sigma\_obs}    & $10^{-18}$\,erg\,s$^{-1}$\,cm$^{-2}$\,arcsec$^{-2}$ & Observed--frame $1\sigma$ Ly$\alpha$ surface--brightness sensitivity per IFU exposure. \\
\texttt{r\_iso}             & kpc                                 & Isophotal radius where Ly$\alpha$ intensity in a circular aperture falls below the \\
& & sky background level ($\sigma$ sky). \\
\texttt{r\_s}               & kpc                                    & Best fit scale length of the exponential component. \\
\texttt{r\_s\_err}          & kpc                                    & $1\sigma$ error on \texttt{r\_s}. \\
\texttt{area\_iso\_2sigma}  & arcsec$^{2}$                           & Isophotal sky area based on 2$\sigma$ sky + continuum-subtracted, background. \\
\texttt{area\_r\_iso\_circ} & arcsec$^{2}$                           & Isophotal sky area based on circular aperture defined by \texttt{r\_iso}.\\
\texttt{logL\_lya}        & $\log_{10}$(erg\,s$^{-1}$)               & Ly$\alpha$ luminosity converted from \texttt{flux\_lya}. \\
\texttt{logL\_lya\_err}   & $\log_{10}$(erg\,s$^{-1}$)               & $1\sigma$ error on \texttt{logL\_lya}. \\
\texttt{flux\_lya}          & erg\,s$^{-1}$\,cm$^{-2}$               & Ly$\alpha$ flux in the circular aperture defined by \texttt{r\_iso}. \\
\texttt{flux\_lya\_err}     & erg\,s$^{-1}$\,cm$^{-2}$               & $1\sigma$ error on \texttt{flux\_lya}. \\
\texttt{gmag}               & mag                                    & $g$-band AB magnitude as measured in the HETDEX PSF spectrum. \\
\texttt{HSC-r\_mag}         & mag                                    & HSC-$r$ AB magnitude in a $2''$ aperture. \\
\texttt{HSC-r\_mag\_err}    & mag                                    & $1\sigma$ error on HSC magnitude. \\
\texttt{combined\_eqw\_rest\_lya} & \AA                              & Rest-frame Ly$\alpha$ equivalent width from \citet{davis2023}. \\
\texttt{flag\_resolved}     & 1=resolved, 0=point-source       & Final decision whether source is resolved based on criteria described in Section\,\ref{sec:resolved}.\\
\texttt{chi2\_ext\_reduced} & \textemdash{}                          & Reduced $\chi^{2}$ of the extended, two-component (PSF + exponential) fit. \\
\texttt{chi2\_psf\_reduced} & \textemdash{}                          & Reduced $\chi^{2}$ of the PSF-only fit. \\
\texttt{log10\_pF} & \textemdash{} &
$\log_{10} p_F$ statistic from the F-test. A source is accepted as resolved if $\log_{10} p_F < -6$.\\
\texttt{dBIC}               & \textemdash{}                          & Difference in the Bayesian Information Criterion. Resolved if dBIC$<-5$. \\
\texttt{iso\_rel\_err} & \textemdash{}  & Fractional deviation between the measured and model-predicted isophotal radii \\
& & is described in Section\,\ref{sec:resolved}.\\
\texttt{dups\_detectid} & \textemdash{} & List of duplicate HLAN names for repeat observations.\\
\hline
\end{tabular}
\caption{Column information for the HETDEX Lyman-$\alpha$ Nebulae Catalog.
The HLAN Catalog can be accessed online at \url{https://hetdex.org/data-results/}
and in the electronic version of this \href{https://doi.org/10.3847/1538-4357/ae44f3}{paper}. Only the header of this table is shown here to demonstrate its form and content. The column names in the online journal version differ slightly to conform to journal standards.}
\label{tab:main}
\end{table*}

The HETDEX \lya\ Nebulae (HLAN) Catalog, containing the best-fit surface-brightness model parameters and LAN source information, is available online at \url{https://hetdex.org/data-results/} and in the electronic version of this paper. A description of the table columns is given in Table~\ref{tab:main}. The name given to each object follows the nomenclature ``HLAN'' supplemented by a representative \texttt{detectid} value where \texttt{detectid} is the integer ID used by the internal HETDEX data model to represent a HETDEX detection. 

In the HETDEX catalog, every point-source emission-line detection is assigned a unique \texttt{detectid}. For spatially resolved objects, such as LANs, multiple \texttt{detectids} can represent a single source -- either as a result of multiple point-source detections at the same wavelength but different spatial positions, or multiple spectral emission-line detections at different wavelengths but coincident spatial locations. During catalog generation, described in detail in \citet{emc2023}, a series of friends-of-friends detection groupings in both 2D and 3D space is applied together with additional logic to reduce multiple detections to a single source. This consolidation is performed on a per-observation basis. Ultimately, a single representative \texttt{detectid} is retained for each LAN\null. This ``selected'' \texttt{detectid} often corresponds to the brightest continuum detection.

If multiple HETDEX observations (labeled \texttt{shotid} in the catalog) cover the same source, this may result in multiple HETDEX detections for the same source and ultimately multiple LAN entries in the HLAN Catalog. We identify HLAN objects with duplicate observations and give their \texttt{detectid} matches in column \texttt{dups\_detectid}. 

As stated above, each surface-brightness model is allowed to raster its location of peak intensity ($\theta=0$) $\pm5\arcsec$ around the source's reported coordinates in HETDEX\null.  The computed centroids are provided in the catalog.  In many cases, these positions agree well with the HETDEX detectid coordinates, but subtle differences in the source-grouping code or redshift assignment do sometimes produce offsets. We store the best-fit coordinates in the catalog in the columns \texttt{ra} and \texttt{dec}. The median separation between this best-fit peak in emission and the representative detection of the source is 0\farcs51 with an interquartile range of 0\farcs35–0\farcs74.

From the initial sample of \nlaefull\ LAE candidates, \nlan\ sources are best fit by a two-component surface-brightness model consisting of a point-source core plus an exponential halo whose flux contributes more than 50\,\% of the total \lya\ intensity. To identify these objects in the catalog, select on column \texttt{flag\_resolved}, where 1=PSF+EXP model preferred, and 0=PSF model preferred.

Ultimately, HETDEX is limited in both image quality and sensitivity, and thus, if a source is not identified as extended in this work, it does not mean that it lacks a significant \lya halo component. Indeed, studies \citep{Wisotzki2016, Leclercq2017} at higher spatial resolution and more sensitive surface-brightness limits find that all LAEs have an extended component.

\subsection{Duplicate LANs and Internal Consistency}

We searched for repeat detections of the same source by cross-matching sky positions.  All pairs of detections within 5\arcsec\ were identified and grouped using a connectivity graph, resulting in 1837 distinct duplicate sources in the catalog.  The unique identifiers (\texttt{detectid}) of neighboring detections are stored in the catalog column \texttt{dups\_detectid}, which lists the set of duplicate sources (comma separated) for each entry.

The presence of repeat observations enables an internal check on the robustness of our extended–source classification. Among all duplicate groupings, we find that approximately 60\% of sources receive consistent resolved/unresolved classifications. In regions with exceptionally deep data (reaching $\sim 2\times$\,\sbunits), this agreement increases to about 80\%. This trend reflects the strong dependence of extended-source identification on surface–brightness sensitivity. When stacking HETDEX sources, extended Ly$\alpha$ emission is detected for stacked LAE surface-brightness profiles \citep{LujanNiemeyer22a, LujanNiemeyer22, McKay2025} and it is likely that, given better sensitivity, every HETDEX source is likely extended in \lya emission as is found in \citet{Wisotzki2016}.

To evaluate the internal repeatability of key observables, we compared measurements of the isophotal radius ($r_{\mathrm{iso}}$), exponential scale radius ($r_{\mathrm{s}}$), isophotal area ($A_{\mathrm{iso,2\sigma}}$), Ly$\alpha$ flux ($F_{\mathrm{Ly\alpha}}$), and Ly$\alpha$ luminosity ($L_{\mathrm{Ly\alpha}}$) among all members of each duplicate group.  The median fractional scatter among repeats was $\sim$0.08 for $r_{\mathrm{iso}}$, 0.12 for $r_{\mathrm{s}}$, 0.13 for $A_{\mathrm{iso,2\sigma}}$, and 0.12 for $F_{\mathrm{Ly\alpha}}$, corresponding to a typical variation of 10–16\%.  The derived luminosities show similarly small differences, confirming that photometric and flux calibration consistency across independent HETDEX observations is excellent.

\subsection{Active Galactic Nucleus Hosts}

The internal HETDEX AGN Catalog for HDR5, (public versions are available in \citet{liu2022, liu2025}), contains 11,995 AGN in the redshift interval where the \lya line falls within the VIRUS spectral window ($1.9<z<3.5$) but only 7890 persist after selection cuts and the analysis is cleaned for poor model fits (see Section~\ref{sec:sample} \& \ref{sec:fitquality}).  Thus the parent sample has an incidence of 11.2\% AGN.  AGN identifications are given in column \texttt{source\_type} of the HLAN catalog, and rely on at least one of the following criteria:

\begin{enumerate}
    \item Broad rest-frame ultraviolet emission with ${\rm FWHM}>1000\;{\rm km\,s^{-1}}$,
    \item A prior AGN identification in SDSS spectroscopy,
    \item The simultaneous detection of \lya and a high-ionisation companion line (typically \ion{C}{4}) at a consistent redshift.
\end{enumerate}

In total, 4106 out of 7890 AGN ($52.5\%$ of the AGN sample) are better fit with a two-component (core\,+\,exponential) surface-brightness model than by a single point-source profile. The incidence of AGN in the parent sample is similar to the fraction of AGN in the LAN sample. We find that 12.2\% of LANs are identified as AGN.

The 50\,\% halo-flux threshold adopted in our classification is intentionally conservative.  Visual inspection of several AGN whose best-fit solution is the single-component model still reveals low-surface-brightness \lya emission beyond the seeing disk. For these brightest AGN, the strong central core dominates the emission, and the sources are categorized best by the one-component model. Two examples can be seen at the bottom of Figure~\ref{fig:ex2}. This leads to a decrease in the extended fraction among the most luminous AGN, as is discussed later in Section~\ref{sec:frac_ext}. A more detailed decomposition of nuclear and extended contributions for the entire AGN sample is deferred to future work.

\section{Results}\label{sec:results}

\begin{figure*}[th]
\includegraphics[width=\textwidth]{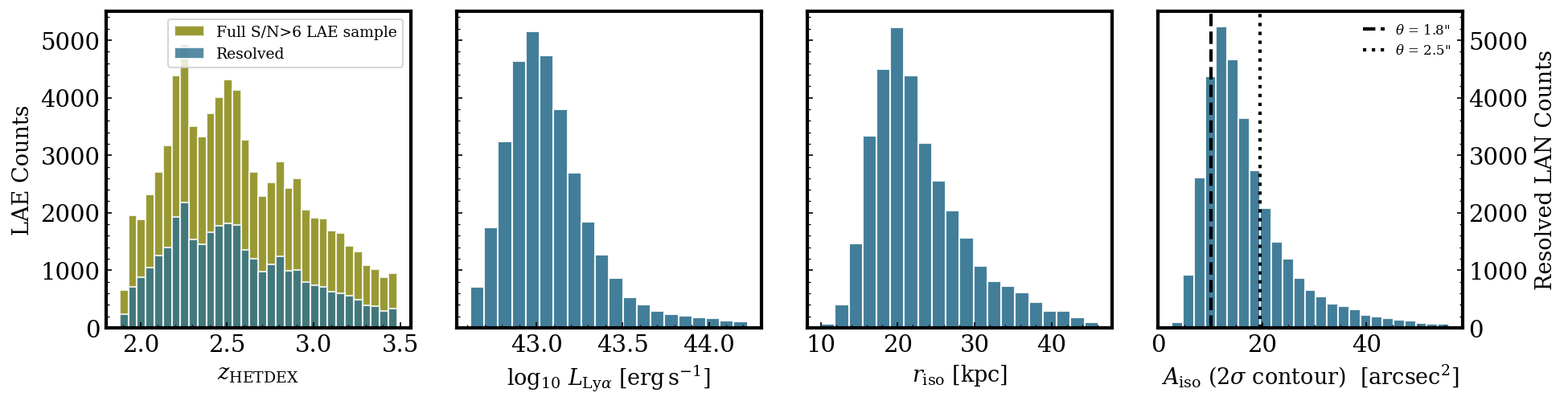}
\caption{\textbf{Left:} redshift distribution; the green histogram shows the full S/N\,$>$\,6 LAE sample, while the blue histogram shows the resolved LAN subset. \textbf{Middle/right:} distributions for the resolved subset of (from left to right) $\log_{10} L_{\mathrm{Ly}\alpha}$ (erg\,s$^{-1}$), isophotal radius $r_{\rm iso}$ (kpc), and isophotal area $A_{\rm iso}$ measured at the $2\sigma$ surface-brightness contour (arcsec$^2$). All panels share a common $y$-axis (counts); tick labels are shown on the far left and right for readability. In the $A_{\rm iso}$ panel, vertical reference lines mark $A=\pi\theta^2$ for seeing FWHM $\theta=1.8''$ (dashed) and $2.5''$ (dotted).}
\label{fig:lan_hists}
\end{figure*}

The resolved LAN sample contains \nlan\ sources. The redshift distributions of the HLAN LANs and the S/N $> 6$ HETDEX LAEs are shown in the left panel of Figure~\ref{fig:lan_hists}. The distributions appear quite similar except for a noticeable difference at higher redshifts ($z>2.8$). This decrease is likely due to the effects of cosmological surface-brightness dimming rather than any evolutionary effect.

The LAN sample spans a range of sizes and luminosities, with their distributions plotted in Figure~\ref{fig:lan_hists}, with a median isophotal radius of $r_{\rm iso}=21.7$\,kpc (16th--84th percentile range: 17.3--29.3\,kpc), and 152 systems exceeding 50\,kpc. For comparison, the exponential scale lengths have a median value of $r_s = 11.8$\,kpc (8.55--16.9\,kpc). A comparison of these two radii is shown in the left panel of Figure~\ref{fig:radius_area_comps}. This demonstrates that the measured isophotal radius is governed primarily by the nebula’s total \lya\ flux rather than an intrinsic size--luminosity relation. Because $r_{\rm iso}$ is defined at a fixed $1\sigma_{\rm sky}$ surface--brightness level, galaxies with low integrated flux fall below the detection threshold at smaller radii, whereas more luminous systems remain detectable to larger distances. The fitted exponential scale length $r_s$ is less sensitive to this surface--brightness limit and therefore provides a more robust tracer of the intrinsic physical extent. Any interpretation of $r_{\rm iso}$ must therefore account for these surface--brightness selection effects, which are discussed further in Section~\ref{sec:sizeluminosity}.

Their Ly$\alpha$ luminosities span $\log_{10}L_{\mathrm{Ly}\alpha}=42.34$--$44.82$\,erg\,s$^{-1}$, with a median of 43.15 and a 16th--84th percentile range of 42.92--43.45. The median Ly$\alpha$ flux is $19.9\times10^{-17}$\,erg\,s$^{-1}$\,cm$^{-2}$ (12.6--38.8). The isophotal area measured at the $2\sigma$ surface--brightness contour has a median of $14.8$\,arcsec$^2$ (9.94--24.4\,arcsec$^2$), while the circularized area from $r_{\mathrm{iso}}$ is systematically larger, with a median of $21.8$\,arcsec$^2$ (16.6--32.8\,arcsec$^2$). 

As a robustness check, we repeated the analysis for a representative test sample using fixed-width pseudo-narrowband images. We find that a wide fixed window ($\sigma = 8$) yields results consistent with our fiducial variable-width approach, while an overly narrow choice ($\sigma = 2$) systematically suppresses extended Ly$\alpha$ emission and increases classification changes for fainter halos that lie on the decision boundary between core and halo dominated.

\subsection{Isophotal Area Comparison}
\label{sec:area_comparison}

\begin{figure*}[t]
  \centering
  \includegraphics[width=\textwidth]{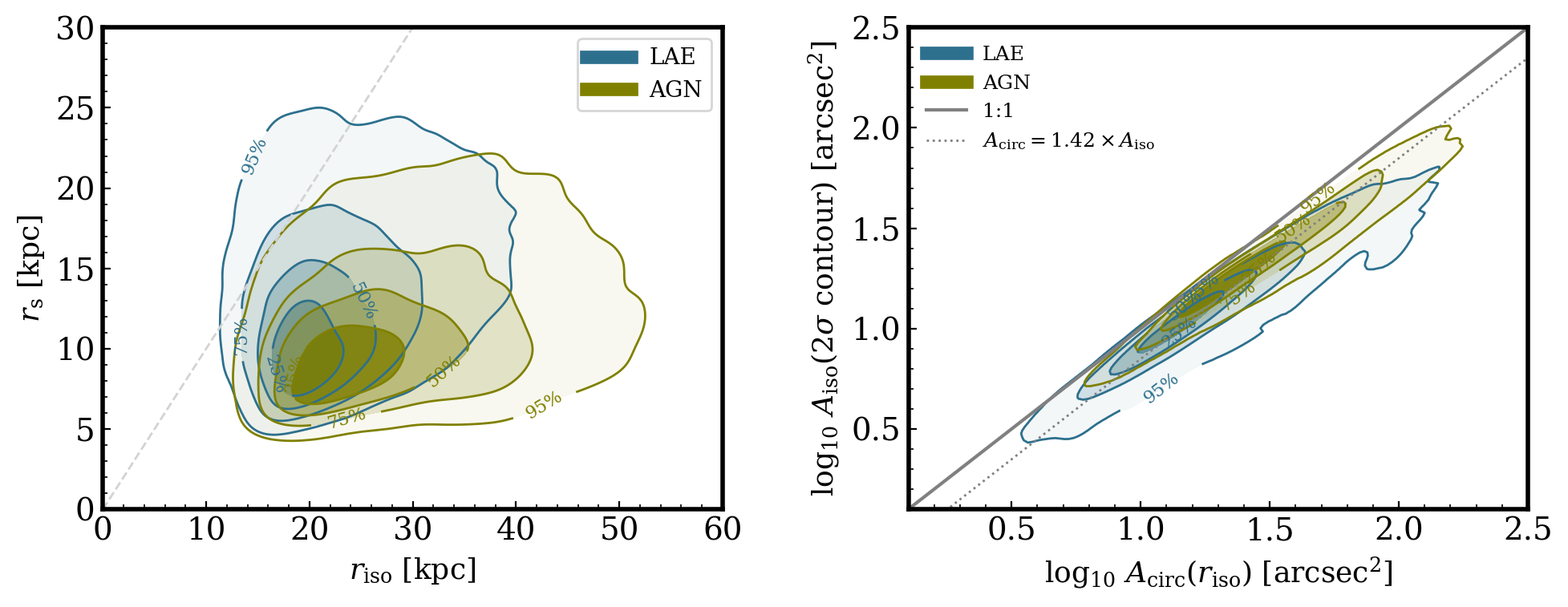}
  \caption{
  Structural comparisons among HETDEX \lya\ Nebulae (LANs). 
  \textbf{(a)}~Exponential scale length, $r_s$, versus isophotal radius, $r_{\mathrm{iso}}$, measured at the 1$\sigma$ surface--brightness limit. 
  Blue and green contours denote LAEs and AGN, respectively; the dashed line marks $r_s = r_{\mathrm{iso}}$. 
  Most sources lie below this relation, confirming that detectable \lya\ emission generally extends beyond the fitted exponential scale length. 
  \textbf{(b)}~Comparison between the measured isophotal area, $A_{\mathrm{iso}}$ (2$\sigma$ \lya\ contour), and the circularized area derived from the 1$\sigma$ isophotal radius, $A_{\mathrm{circ}} = \pi r_{\mathrm{iso}}^2$. 
  The solid line shows the one-to-one relation, and the dashed line marks $A_{\mathrm{circ}} = 1.5\,A_{\mathrm{iso}}$, close to the sample median. 
  Both panels demonstrate that the circularized radius provides a reliable approximation to the isophotal extent of the \lya\ emission, while the exponential component typically underestimates the full halo size. 
  The median ratio $A_{\mathrm{circ}}/A_{\mathrm{iso}}=1.42$ corresponds to radii that are $\simeq$20\% larger, confirming that the circularized $r_{\mathrm{iso}}$ closely reproduces the measured isophotal extent.}
  \label{fig:radius_area_comps}
\end{figure*}

To assess the correspondence between the effective isophotal radius ($r_{\rm iso}$, measured at 1$\sigma$, \texttt{SB\_1sigma\_obs} in the catalog) and the total area enclosed by the 2$\sigma$ surface-brightness contour (column \texttt{area\_iso\_2sigma}), we compare the circularized area inferred from $r_{\mathrm{iso}}$ (\(A_{\mathrm{circ}}=\pi r_{\mathrm{iso}}^2\), available in column \texttt{area\_r\_iso\_circ}) with the directly measured isophotal area, $A_{\mathrm{iso,2\sigma}}$, in the right panel of Figure~\ref{fig:radius_area_comps}. The majority of sources cluster around the one-to-one relation, but with a systematic offset, indicating that the circularized radius provides a robust proxy for the total isophotal extent, albeit with moderate scatter toward smaller $A_{\mathrm{iso}}$ values at higher surface-brightness thresholds. The distribution is centered above unity with ${\rm median}(R)=1.42$ and an interquartile range of $[1.33,\,1.75]$, implying that the circularized area inferred from $r_{\mathrm{iso}}$ typically exceeds the 2$\sigma$ isophotal area by approximately 50\%.

A small minority of systems show $R<1$ (\(R_{\min}=0.32\)), consistent with cases where the 2$\sigma$ isophote extends asymmetrically beyond the circularized aperture—such as filamentary morphologies, centroid offsets, or blended structures. Conversely, large $R$ outliers likely occur when circularization overestimates the extent of elongated or patchy halos, when $r_{\mathrm{iso}}$ measured at the 1$\sigma$ level captures low S/N wings absent at 2$\sigma$, or from minor measurement systematics (e.g., segmentation or masking effects, deblending errors, or redshift-dependent $\,\mathrm{kpc}/\arcsec\,$ conversions for a small subset of sources).

Overall, this comparison confirms that the circularized radius measurement provides a reliable and easily reproducible proxy for total Ly$\alpha$ extent, with predictable biases that can be statistically characterized for large samples.

%The isophotal area has a median of $25.0$\,arcsec$^2$ (18.9--85.7~arcsec$^2$), and the \lya luminosities range from log values of $42.4$ to $44.9$ erg~s$^{-1}$, with a median value of 43.2 and a typical spread of $42.9$--$44.1$. The distribution of isophotal area can be seen in the right panel of Figure~\ref{fig:lan_hists}.

\subsection{Optical Counterparts}

\begin{figure*}[t]
  \centering
  \includegraphics[width=\textwidth]{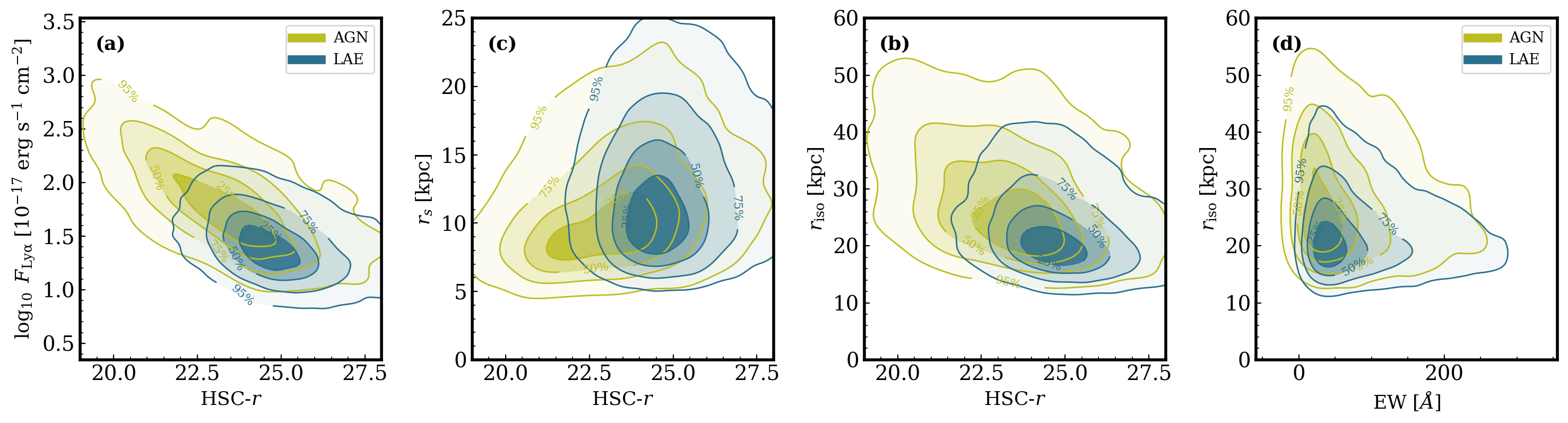}
  \caption{Optical magnitudes of HETDEX \lya\ Nebulae (LANs) with identified HSC counterparts. 
  From left to right: 
  \textbf{(a)}~\lya\ flux versus HSC--$r$ aperture magnitude; 
  \textbf{(b)}~exponential scale length, $r_s$, as a function of HSC--$r$; 
  \textbf{(c)}~isophotal radius, $r_{\mathrm{iso}}$, as a function of HSC--$r$; and 
  \textbf{(d)}~isophotal radius versus rest-frame \lya\ equivalent width (EW). 
  Blue and green contours correspond to LAEs and AGN, respectively. 
  LAEs are typically $\sim$1--1.5\,mag fainter in the optical continuum than AGN yet span comparable \lya\ fluxes. 
  Both $r_s$ and $r_{\mathrm{iso}}$ show weak dependence on continuum brightness, though the most extended haloes 
  ($r_s\gtrsim20$\,kpc; $r_{\mathrm{iso}}\gtrsim40$\,kpc) are preferentially associated with the optically brightest AGN.}
  \label{fig:optical_counterparts}
\end{figure*}

Nearly the entire HETDEX spring field has been imaged in the $r$-band to $r\sim26.1$ with Hyper Suprime-Cam \citep[HSC;][]{davis2023}; similarly, the other HETDEX fields considered in this paper have $grizy$ coverage down to $r\sim 26$ from Subaru Strategic Program (HSC-SSP; \citealt{HSC-SSP-Survey}). These data enable a direct comparison between the continuum counterparts of our \lya detections and their emission-line properties. Aperture magnitudes on their $r$-band continuum images were measured in a series of circular apertures centered on the LAN centroid position and grown until the enclosed optical flux converged. These are found in the catalog in column \texttt{HSC-r\_mag} if available.  Where no counterpart is present at the $3\sigma$ level ($r\gtrsim29$) we fix an upper limit at $r = 30$ in the catalog, but do not plot these values.

Figure~\ref{fig:optical_counterparts} summarizes the relation between continuum brightness, \lya\ flux, and the extent of the \lya\ emission for HETDEX LANs.
Panels (a)--(c) show that LAEs (blue) and AGN (green) span similar \lya\ fluxes and halo sizes despite a systematic offset in continuum brightness, with LAEs typically $\sim$1--1.5\,mag fainter in HSC--$r$. In (a) we see that \lya\ flux generally scales with continuum magnitude. In (b) and (c), the exponential scale length $r_s$ and isophotal radius $r_{\mathrm{iso}}$ display little dependence on continuum magnitude, suggesting that the spatial extent of the \lya\ emission is only weakly tied to the stellar or AGN continuum luminosity.
However, the largest haloes ($r_s\gtrsim20$\,kpc; $r_{\mathrm{iso}}\gtrsim40$\,kpc) are preferentially found among optically bright AGN, consistent with enhanced \lya\ scattering or illumination in AGN-driven ionization cones.

Panel~(d) relates the physical halo size to the rest--frame equivalent width (EW). Generally, the larger $r_{\mathrm{iso}}$ LANs have lower EWs as they tend to be associated with brighter continuum counterparts (as seen in Panel~(c)). This trend supports a picture in which large-scale, low-surface-brightness \lya\ emission contributes substantially to the total line-flux, boosting EW measurements for galaxies embedded in more diffuse, resonantly scattered haloes which are not associated with bright-continuum counterparts.

Together, these trends underscore two key points and echo the recent results of \citet{Li2024}: (i) the majority of LAEs in HETDEX are faint or even undetected in deep $r$-band imaging, reinforcing the power of blind IFU surveys to uncover low-mass, continuum-faint galaxies; and (ii) AGNs contribute disproportionately to the bright-continuum, low-EW tail of the LAE distribution but represent only a minor fraction ($\sim12$\%) of the overall population of \lya emitters.

\subsection{How Many LAEs are Extended?}
\label{sec:frac_ext}

\begin{figure*}[ht]
  \centering
  \includegraphics[width=\linewidth]{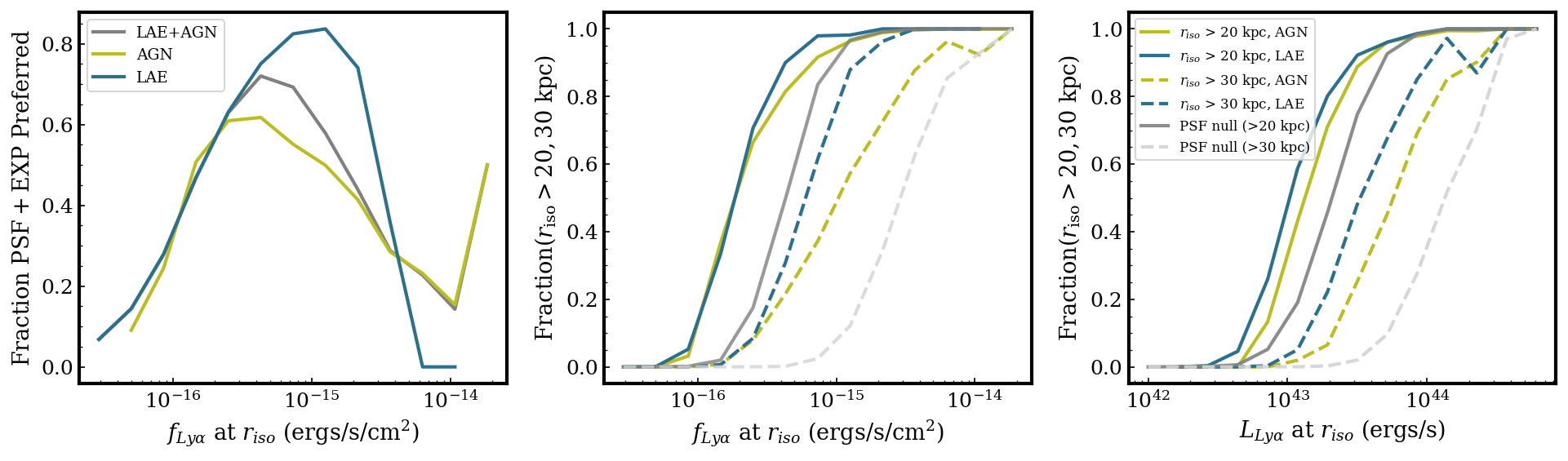}

  \caption{Incidence of spatially extended \lya\ nebulae in the S/N$>6$ parent LAE sample. Left: fraction of sources that prefer the two–component (PSF+exponential envelope) model over a single PSF–only fit. At low ($f_{Ly\alpha}<10\times$\,\fluxunit) and high ($f_{Ly\alpha}>100\times$\,\fluxunit) line-flux values, both AGN and LAE have higher incidences of point-source preferred models. Middle: fraction of sources whose isophotal radius exceeds either $20$\,kpc (solid) or $30$\,kpc (dashed) as a function of the \lya\ flux measured at $r_{\mathrm{iso}}$. Right: same as the middle panel, but binned by the \lya\ luminosity enclosed within $r_{\mathrm{iso}}$. Blue and green curves denote galaxies classified as LAEs and optically identified AGN \citep{liu2025}, respectively. Gray lines show the PSF-only null hypothesis computed for each object from its own seeing, surface-brightness limit, and redshift. The observed incidence of extended emission rises steeply with increasing \lya\ surface brightness or luminosity, well above the expectation for unresolved sources.}
  \label{fig:frac_rext}
\end{figure*}

Figure~\ref{fig:frac_rext} quantifies, for the first time in an wide-area, untargeted IFU survey, how frequently luminous LAEs at $z\simeq 2$--3 at $1\sigma$ \lya sensitivities of $\sim2-5\times$\sbunits\ display spatially extended \lya emission. We first consider the fraction of sources that prefer the two-component (PSF+exponential) model fit. For a series of \lya flux bins, we calculate the fraction of sources that satisfy the extended-source criteria described in Section~\ref{sec:resolved}, divided by the total number of sources in each flux bin. The ensemble sample is shown in solid gray, while those classified as LAEs and AGNs are shown by the blue and green curves, respectively. At both low ($f_{\mathrm{Ly}\alpha}<10\times$\,\fluxunit) and high ($f_{\mathrm{Ly}\alpha}>100\times$\,\fluxunit) line-flux values, both AGN and LAEs exhibit higher incidences of point-source–preferred models.

In the lower flux regime, the data lack sufficient S/N to provide resolved fits to the data, so we cannot conclude that these LAEs truly lack extended \lya halos. In \citet{McKay2025}, we show that stacks of HETDEX LAEs with good image quality reveal that extended emission is still present on average.

At higher fluxes, visual inspection shows that sources best described by the PSF-only model—without a significant exponential component—tend to have bright optical counterparts. In the AGN sample, there are spectral signatures of AGN hosts (broad emission, AGN emission line pairs, bright-continuum optical counterparts). Two examples of bright AGN with strong \lya\ emission but little extended structure are shown in the bottom two panels of Figure~\ref{fig:ex2}. In the LAE sample, the sources that prefer a one-component model appear similar to the AGN cases: they have luminous optical counterparts but lack clear AGN features in the HETDEX spectra. It is plausible that many of these are, in fact, narrow-line AGN with other emission lines, such as \CIV\ lying outside the HETDEX spectral window. We note that, upon visual inspection, even among the brightest systems, some extended emission remains evident, but it contributes less than 50\% of the total \lya\ flux. Consequently, the two-component model offers no statistical preference over the simpler PSF-only fit.  This behavior reflects the fact that our classification is based on statistical model comparison rather than on residual flux after PSF subtraction, and is therefore sensitive to the dominance of the central source at high luminosities, particularly because the extended component must contribute a substantial fraction of the total Ly$\alpha$ flux to be statistically favored.

In the middle panel, we consider the fraction of objects whose isophotal radius exceeds either $20$ or $30$\,kpc --- thresholds that bracket the classical definitions of ``medium'' and ``large'' LANs --- as a function of both the \lya flux, in the middle panel, and intrinsic luminosity in the right panel. These are aperture values determined by the circular aperture defined by $r_{\rm iso}$. In these panels, we separate those defined as AGN (green) from those defined as LAE (blue). The fractions are determined for the full parent sample of \nlaegood\ LAEs.

The extent of $r_{\rm iso}$ is dependent on both line-flux and the surface brightness of the data, and will increase even in the case of a PSF-only surface-brightness model. Even if the sample consisted entirely of unresolved sources, a trend would be seen with flux. We consider this PSF-only null hypothesis in the gray curves in the middle and right panels. We estimate the apparent isophotal radius expected for an unresolved source by convolving a point source with the seeing profile described by a circular Moffat function and evaluating where the surface brightness falls below the 1$\sigma$ detection limit. The calculation is performed for each object using its own FWHM, surface-brightness limit, and redshift to account for variations in angular scale and depth. We then determine the fraction of such point-source models that would exceed 20\,kpc (solid gray curve) or 30\,kpc (dashed gray curve) at the survey isophote.

At low fluxes, the observed fractions closely follow the null prediction, indicating that nearly all sources are unresolved. As flux increases, however, both LAE and AGN populations rise well above the gray curves, implying that the measured sizes cannot be explained by PSF broadening alone. The excess fraction above the null model, therefore represents genuine spatially extended \lya\ emission. LAEs, particularly those with $r_{\rm iso}>30$\,kpc (dashed blue curve), show a systematically higher incidence of extended radii than AGN (dashed green curve) at fixed flux, suggesting that their halos are more prevalent or more diffuse for a given brightness.

The right panel presents the same metric as a function of intrinsic \lya\ luminosity. The trends mirror those in the middle panel but remove the influence of distance on the observed flux. The offset between the colored and gray curves again indicates that extended emission is common beyond what would be expected for unresolved sources and that LAEs are more likely to be extended than an AGN-dominated sample. The 50\% incidence threshold for $r_{\mathrm{iso}}>30$ kpc occurs near $L_{\mathrm{Ly}\alpha}\sim3\times10^{43}$ erg s$^{-1}$ for LAEs and slightly higher for AGN at $L_{\mathrm{Ly}\alpha}\sim3\times10^{43}$ erg s$^{-1}$, consistent with a luminosity-dependent transition in which nearly all systems above $10^{44}$ erg s$^{-1}$ host large-scale \lya\ halos. 

Together, the middle and right panels demonstrate that the frequency and extent of extended \lya\ emission increase steeply with both flux and luminosity, and that LAEs remain more spatially extended than AGN over the full dynamic range. These results show that extended LANs are common but not ubiquitous. Their prevalence depends on both intrinsic Ly$\alpha$ output and the nature of the power source, with the systematic offset between AGN and star-forming LAEs suggesting fundamental differences in the mechanisms that excite Ly$\alpha$ emission.

\subsection{The Ly$\alpha$ Halo Size-Luminosity Relation}
\label{sec:sizeluminosity}

\begin{figure*}
    \centering
    \includegraphics[width=\linewidth]{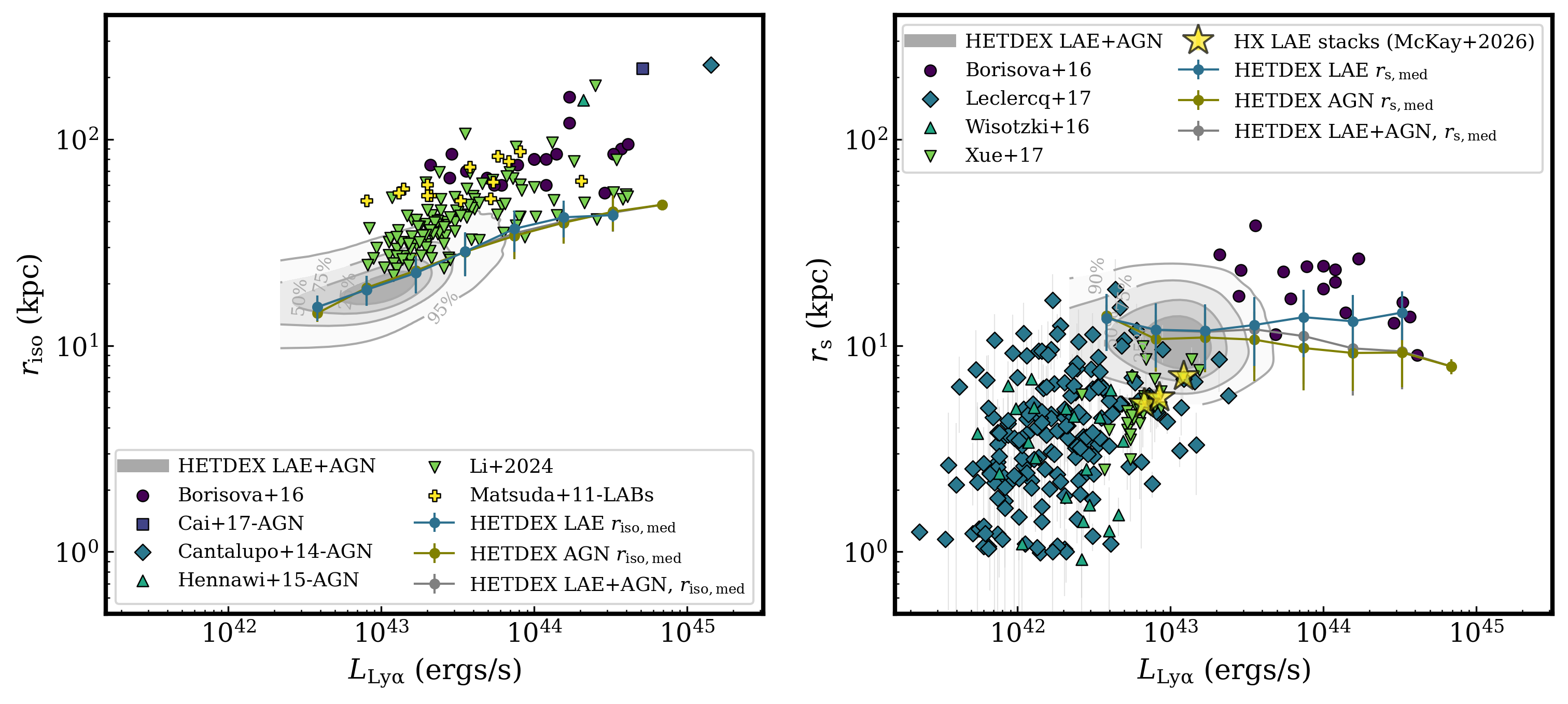}
    \caption{ Radial extent versus \lya\ luminosity measurements based on different size estimators. Left: isophotal radius ($r_{\mathrm{iso}}$) versus Ly$\alpha$ luminosity ($L_{\mathrm{Ly\alpha}}$) for HETDEX LAEs and AGN (gray contours), compared with previous studies \citep{matsuda11, cantalupo14, hennawi15, borisova16,cai17, Li2024} of extended Ly$\alpha$ emission (colored symbols).  Median HETDEX sizes in bins of luminosity are shown separately for LAEs (blue), AGN (green), and the combined sample (gray).  Right: exponential scale length ($r_{\mathrm{s}}$) versus $L_{\mathrm{Ly\alpha}}$ for the same sample from \citet{borisova16}, and additional measurements of \lya\ halos from \citet{Leclercq2017}, \citet{Wisotzki2016}, and HETDEX LAE stacks from \citet{McKay2025}. gray shaded regions in both panels show kernel-density estimates (KDEs) enclosing 25–95\% of the HETDEX LAE+AGN population.}
    \label{fig:r_vs_L}
\end{figure*}

To explore the relationship between Ly$\alpha$ halo size and luminosity, we compare HETDEX LAEs to a compilation of literature sources spanning diverse environments and measurement techniques. Figure~\ref{fig:r_vs_L} shows the distribution of isophotal radii ($r_{\mathrm{iso}}$, left panel) and exponential scale radii ($r_{\mathrm{s}}$, right panel) as functions of $L_{\mathrm{Ly\alpha}}$. In the left panel, we include isophotal measurements from LABs compiled from \citet{borisova16}, \citet{matsuda11}, \citet{cai17}, \citet{cantalupo14}, \citet{hennawi15} and \citet{Li2024}. We apply a correction factor of 0.5 to approximate isophotal radii from full angular extent measurements and note that these will naturally be dependent on the observation's surface-brightness sensitivity. The right panel shows exponential scale lengths from the LAB sample from \citet{borisova16}, and from the LAH samples from \citet{Wisotzki2016}, \citet{Leclercq2017}, and \citet{xue17}, alongside HETDEX LAE stacks \citep{McKay2025}.

To visualize the distribution of the \nlan\ HETDEX sources in the $L_{\mathrm{Ly\alpha}}$–size plane, we computed 2D kernel-density estimates (KDEs) using a Gaussian kernel, evaluated on a uniform logarithmic grid in both luminosity and size. The KDEs are normalized such that their total integral equals unity, and the contours in Figure~\ref{fig:r_vs_L} correspond to the highest-density probability (HDP) regions enclosing 25\%, 50\%, 75\%, and 95\% of the total LAN population. In both panels, HETDEX sources occupy a relatively narrow region in luminosity–size space, with a trend of increasing $r_{\mathrm{iso}}$ with luminosity in the left panel, in line with other sources in the literature. No significant trend in the distribution is seen in the right panel with exponential scale length, $r_{s}$.

The ensemble averages shown for LAEs (blue), AGN (green), and the combined sample (gray) confirm that $r_{\mathrm{iso}}$ increases systematically with $L_{\mathrm{Ly\alpha}}$, consistent with literature data points. In contrast, the exponential scale lengths show little luminosity dependence, implying that the intrinsic halo profiles remain broadly similar across luminosities. This behavior reflects the known sensitivity of isophotal measurements to surface-brightness limits: as emphasized by \citet{steidel2011}, deeper data reveal more extended emission even at fixed luminosity. The scatter in $r_{\mathrm{iso}}$ in HETDEX sources likely arises from a combination of intrinsic variation in the individual source luminosity, redshift, and \lya surface-brightness distribution, as well as measurement differences in sensitivity and image quality. For example, much of the observed scatter can likely be mitigated by adopting a universal surface-brightness threshold that corrects for cosmological dimming, following the approach of \citet{battaia2023}.

Across the combined HETDEX LAE+AGN sample, the median exponential scale length is $r_{\mathrm{s}} = 11.6 \pm 1.9$~kpc. Separating by source type, LAEs exhibit larger median scale lengths ($12.9 \pm 1.5$~kpc) than AGN ($10.3 \pm 1.4$~kpc), revealing differences in emission related to AGN and star-forming regions. Future work will relate HETDEX LAN profiles to galaxy properties in greater depth.

\subsection{Resolved versus PSF Flux Comparisons}
\label{sec:resolved_flux}

IFU observations of LAEs with VLT/MUSE show that much of the \lya energy budget lives in this halo component: individual LAEs contain $\sim40$–$90\%$ of their line-flux at radii that exceed the
seeing disk \citep{Wisotzki2016}, with a median halo contribution of $\simeq65\%$ in the MUSE Hubble Ultra-Deep-Field sample \citep{Leclercq2017}. PSF-weighted or circular apertures recover only a fraction of the total \lya\ emission.  In the narrowband LAE sample of \citet{huang2021}, for example, point-source or fixed angular apertures miss $\sim$30\% of the Ly$\alpha$ flux in a $z\sim3.1$ protocluster.

\begin{figure}
  \centering
  \includegraphics[width=\linewidth]{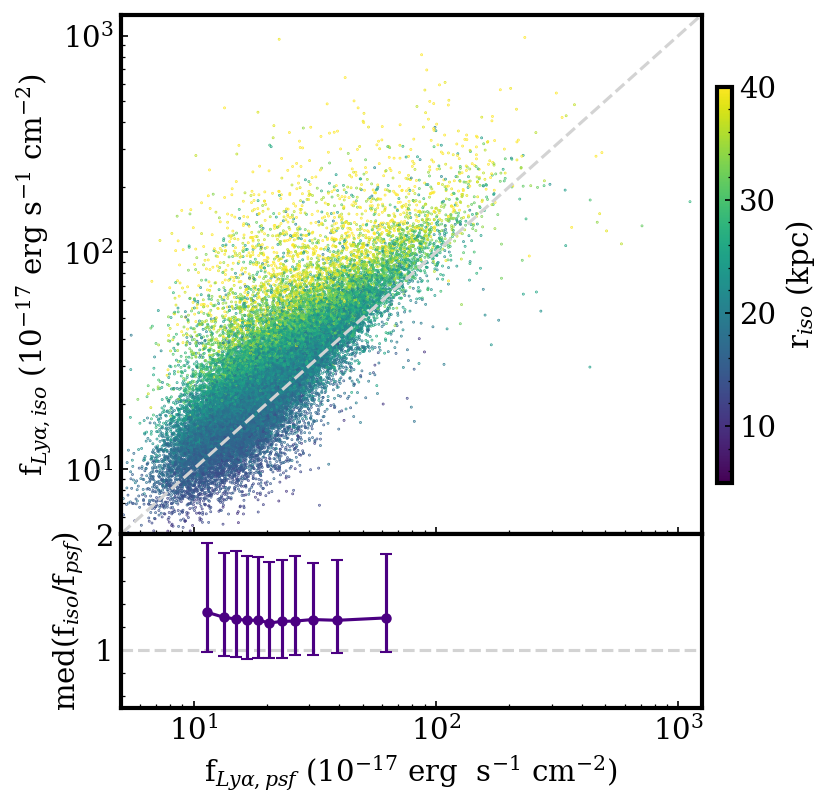}
  \caption{Comparison between the \lya\ fluxes measured in
  isophotal apertures, $F_{r\rm iso}$, and fluxes measured from PSF measurements in the HETDEX reduction pipeline ($F_{\rm psf}$).
  Top: individual LAEs are color coded by their physical
  isophotal radius $r_{\rm iso}$; the dashed light-gray line marks the
  1:1 relation.  Most objects lie above this line, showing that a
  PSF-based extraction systematically misses flux from extended nebular
  emission.  
  Bottom: median flux ratio
  $F_{\rm iso}/F_{\rm psf}$ in 12 adaptive bins. The median ratio is 
 $\simeq1.3$ with little dependence on flux.}
  \label{fig:resolved_flux_comparison}
\end{figure}

Because the HETDEX pipeline extracts a PSF-weighted spectrum by construction, its reported fluxes can be affected by this bias. Figure~\ref{fig:resolved_flux_comparison} quantifies the effect on a source-by-source basis. Nearly every LAE lies above the 1:1 line, confirming that PSF extractions underestimate the integrated Ly$\alpha$
flux. The offset correlates with $r_{\rm iso}$ (shown in color), with the largest objects resulting in larger flux measurement offsets.Across the full dynamic range, the median $f_{\rm Ly\alpha,iso}/f_{\rm Ly\alpha,psf}$ is $\sim1.3$, implying a systematic flux deficit of roughly 30\%. 

This systematic offset will propagate directly into luminosity functions, equivalent-width distributions, star-formation-rate estimates, and any quantity that depends on absolute \lya\ luminosity.

\subsection{HETDEX LANs Observed in Other Surveys}

\begin{table*}[ht]
\centering
\begin{tabular}{lcccc|lcccc}
\hline
\multicolumn{5}{c|}{MAMMOTH} & \multicolumn{5}{c}{This Work (HETDEX)} \\
Name & RA & Dec & Area & log $L$ & HETDEX Name & RA & Dec & Area & log $L$ \\
 & (deg) & (deg) & (arcsec$^2$) & (erg s$^{-1}$) &  & (deg) & (deg) & (arcsec$^2$) & (erg s$^{-1}$) \\
\hline
MLAN43 & 32.1770 &  0.6582 & 41.6 & 43.49 & HLAN4016274387 & 32.1769 &  0.6582 & 39.38 & 43.65 \\
MLAN78 & 32.7873 &  0.7366 & 26.1 & 43.34 & HLAN4015512497 & 32.7870 &  0.7367 & 21.50 & 43.19 \\
MLAN60 & 32.9132 &  1.0364 & 32.2 & 43.36 & HLAN3010736450 & 32.9131 &  1.0361 & 36.06 & 43.43 \\
MLAN85 & 35.5727 & -2.1925 & 24.3 & 43.05 & HLAN3009861289 & 35.5726 & -2.1925 & 13.94 & 43.08 \\
\hline
\end{tabular}
\caption{Comparison of LABs Matched between Our Sample and MAMMOTH. Note: coordinates, solid angle, and Ly$\alpha$ luminosity from both studies are listed. Area is measured from isophotal contours drawn at the varying 2$\sigma$ surface-brightness (column \texttt{area\_iso\_2sigma}) limit of the HETDEX observation.}
\label{tab:mammoth_matches}
\end{table*}

A comparison of our catalog to the recently released LAN sample from the MAMMOTH-Subaru survey \citep{Li2024} reveals four LANs common to both catalogs. This survey reaches similar \lya\ surface-brightness limits to HETDEX, with a typical $2\sigma$ \lya\ surface brightness of $5-10\times10^{-18}$\,erg\,s$^{-1}$\,cm$^{-2}$\,arcsec$^{-2}$. As with HETDEX, the surface brightness varies from field to field. Table~\ref{tab:mammoth_matches} summarizes the comparison, including positions, isophotal areas, and Ly$\alpha$ luminosities from both studies. The matched LANs between our sample and MAMMOTH agree well in position, with coordinate offsets always less than 1\farcs3  (median offset: 0\farcs6). The \lya luminosities (measured within the encompassing isophotal areas) are also consistent, differing by less than 0.16~dex (median $\Delta \log L = 0.05$ dex), well within typical systematic uncertainties from differing measurement apertures and surface-brightness limits. In both cases, the isophotal area comes from the area enclosed in the 2$\sigma$ isophotal boundary (\texttt{area\_iso\_2sigma} in the HETDEX catalog). Overall, the HETDEX and MAMMOTH measurements agree quite well — typical area differences are around 15\%, consistent with expected variations from slightly different surface-brightness thresholds and segmentation methods.

\subsection{Radio Counterparts}

Radio observations of the HETDEX Spring field from the Low Frequency Array (LOFAR; \citealt{lofar2013}) are fully included in the second data release from the LOFAR Two-Metre Sky Survey (LOTSS; \citealt{ lofardr2}). HETDEX counterparts of LOTSS sources were previously identified in \citet{debski2025} from an earlier HETDEX data release, but here we are able to consider radio counterpart fractions obtained from the LOFAR DR2 catalog \citep{lofardr2_cat} for the full parent sample of HETDEX LAEs and compare this to the LAN radio counterpart fraction.

In the full parent sample of LAEs with S/N$>6$, 51,570 LAEs are in the HETDEX Spring field, 1523 (2.9\%) have radio counterparts. Of these LAEs, roughly half the sample (979/1523; 64\%) prefer an extended emission model. Inspection of those LAEs with radio emission that are best represented by a point-source model are generally AGN (296/596) or LAEs with lower S/N. Half of the LAEs in this sample are at S/N$<7$ and lack sufficient signal to detect extended emission.

\begin{figure*}
    \centering
    \includegraphics[width=\linewidth]{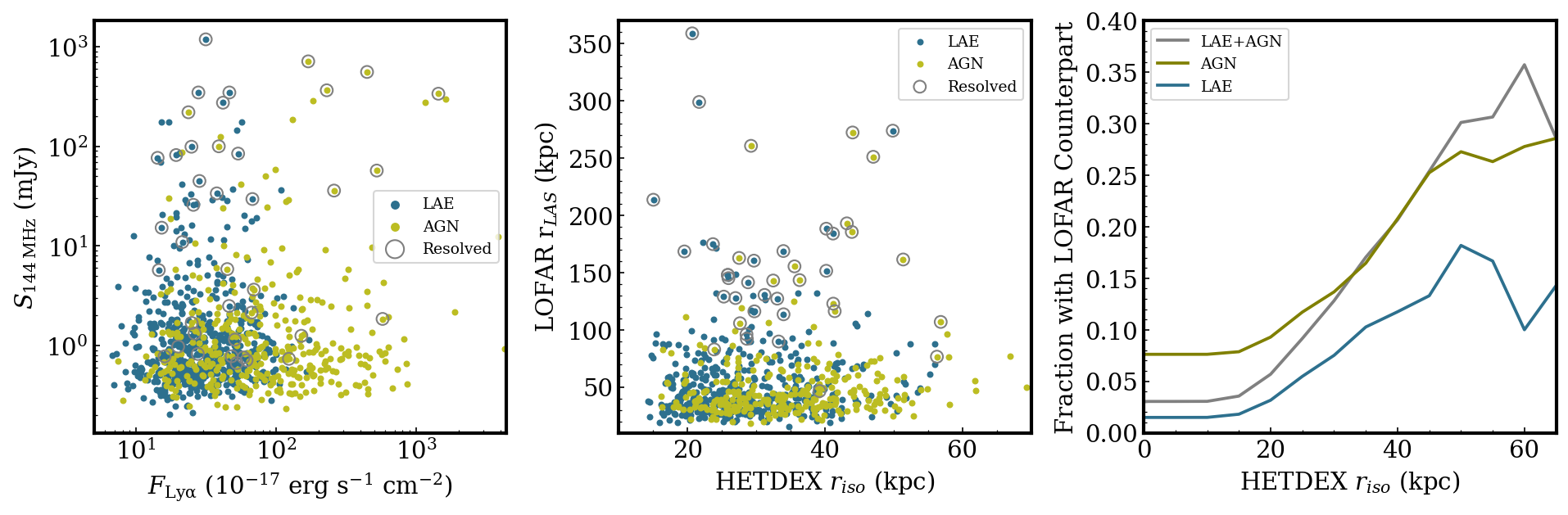}
    \caption{ Left: Ly$\alpha$ line-flux versus integrated LOFAR 144\,MHz flux density. Blue and green symbols indicate LAE and AGN, respectively; gray open circles highlight     sources with spatially resolved LOFAR emission (\texttt{Resolved}~$=$~True in \citealt{lofardr2}). 
    Middle: comparison between the Ly$\alpha$ isophotal radius measured by HETDEX 
    ($r_{\rm iso}$) and the LOFAR largest angular scale converted to proper kpc ($r_{\rm LAS}$). 
    Right: fraction of HETDEX sources with LOFAR counterparts as a function of $r_{\rm iso}$, 
    shown separately for LAE (blue), AGN (green), and the combined sample (gray). Larger Ly$\alpha$ 
    halos show a higher probability of association with LOFAR emission, particularly among AGN.}
    \label{fig:hetdex-lofar-comps}
\end{figure*}

In total, there are 21,566 LANs in the HLAN Catalog in the HETDEX Spring field. Of these objects, roughly 4.5\% (979/21,566), have radio counterparts within 5\arcsec of the peak LAN emission. Of these LAN-LOFAR sources, 371 are labeled AGN. The sample is biased toward higher isophotal radii and \lya\ luminosities. For example, the LAN-LOFAR sources have a median $r_{\rm iso}=28.9$\,kpc compared to 21.7\,kpc and median $\log L_{Ly\alpha}$ of 43.4 erg s$^{-1}$ compared to 43.1\,erg~s$^{-1}$ for the full LAN sample. The fraction of LANs with radio counterparts increases for larger LANs. For instance, LANs with $r_{\rm iso}>30$\,kpc, 14\% of the LANs have radio counterparts, and at $r_{\rm iso}>50$\,kpc, 35\% of the LANs have radio counterparts. This can be seen in the rightmost panel of Figure~\ref{fig:hetdex-lofar-comps} where the fraction of LANs with radio counterparts increases with $r_{\rm iso}$.

\begin{figure*}[ht]
\includegraphics[width=0.48\textwidth]{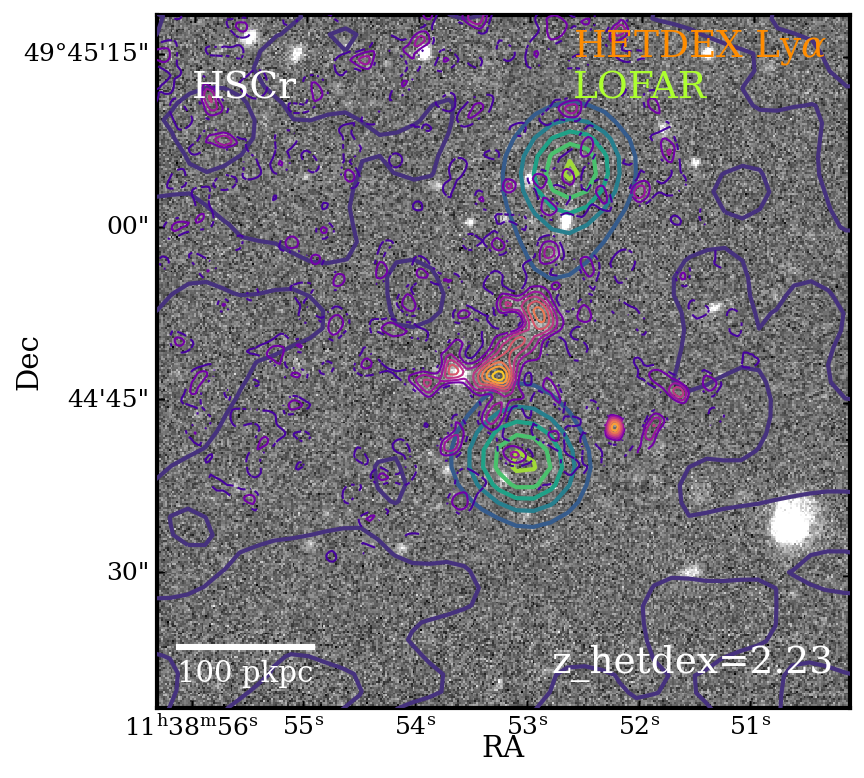} \hfill
\includegraphics[width=0.48\textwidth]{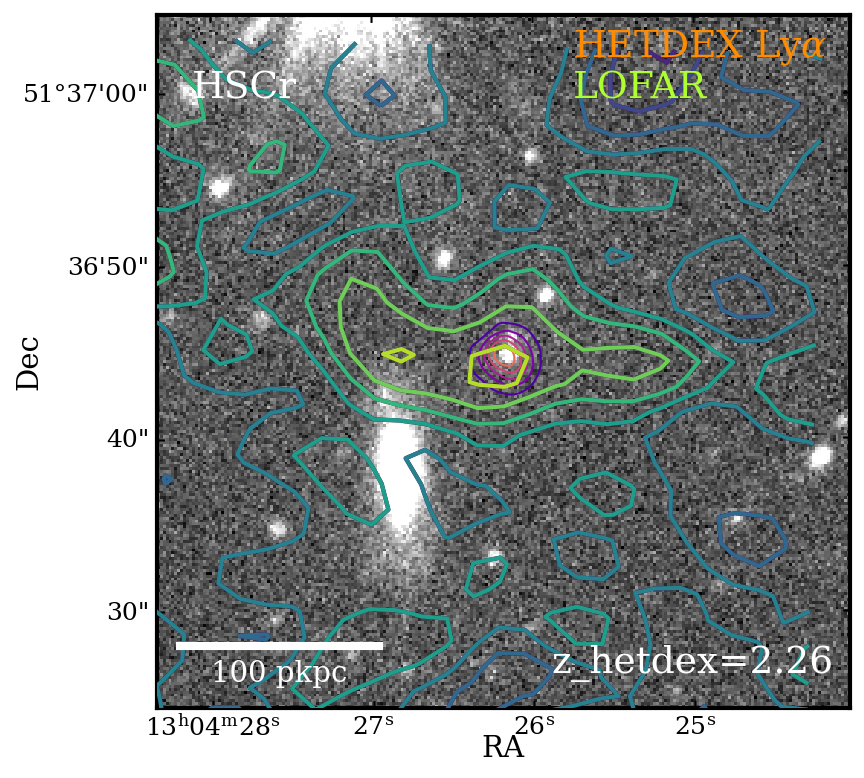}
\caption{ Multi-wavelength overlays for two HETDEX-detected sources exhibiting extended Ly$\alpha$ emission. Left: $60\arcsec$ wode region centered on \texttt{HLAN5002600476} ($z = 2.23$). 
Right: $50\arcsec$ wide region centered on \texttt{HLAN4028050917} ($z = 2.26$), a HETDEX AGN source identified by \lya + CIV emission. The gray-scale background displays $r$-band imaging from HSC-SSP. Overlaid in blue/green are LOFAR 144~MHz radio continuum contours, and in orange/pink are HETDEX \lya\ narrowband emission contours. The LOFAR contours trace extended radio emission, often associated with AGN activity or star-forming regions, while the Ly$\alpha$ contours reveal the spatial extent of ionized gas at high redshift. A 100~kpc scale bar is shown in the lower left of each panel.}
\label{fig:hetdex-lofar-examples}
\end{figure*}

Some extended radio counterparts have HETDEX \lya\ counterparts. In the full S/N$>6$ sample, 66 extended radio sources are found. Nearly two-thirds of these extended radio sources (41/66), are also in the HETDEX LAN sample. For HETDEX LANs with resolved LOFAR detections, we observe a diverse range of radio morphologies. Broadly, they fall into two categories, which we highlight in an example in Figure\,\ref{fig:hetdex-lofar-examples}: some sources, such as the one on the left, exhibit one or two LOFAR lobes flanking the central Ly$\alpha$-emitting region, while others possess extended LOFAR emission that is spatially coincident with the Ly$\alpha$ emission but generally on much larger scales, as seen in the right panel. The morphologies of these \lya-radio sources suggest that many LANs harbour AGN, even those LAEs not previously classified as hosting AGN.

Figure~\ref{fig:hetdex-lofar-comps} compares the Ly$\alpha$ and radio properties of every HETDEX–LOFAR match in the Spring field. Resolved sources generally lie toward the upper envelope of both panels, underscoring that the most extended LANs also tend to host the brightest and most spatially extended radio emission.  This figure highlights the size and flux selection effects discussed above and provides motivation for the morphological categories shown in Figure~\ref{fig:hetdex-lofar-examples}.

\section{Summary}

This study measures the extended surface-brightness profiles of an emission-line–selected sample of Ly$\alpha$-emitting galaxies drawn from the Hobby–Eberly Telescope Dark Energy Experiment (HETDEX). We model \nlaegood\ LAEs as a point-source component with an optional exponential envelope. From this parent sample, we identify \nlan\ Ly$\alpha$ Nebulae (LANs) that are better described by the core+exponential envelope model, representing nearly half (47.5\%) of all LAEs in the parent sample. We also provide area measurements based on 2$\sigma$ isophotal contours, as is common in LAB studies, and an effective isophotal radius measured in circular annuli. While this value trends with \lya flux and luminosity, we show the exponential scale length of the sample is not \lya luminosity dependent

Roughly 12\% of LANs coincide with AGN identified in the HETDEX AGN catalogs of \citet{liu2022, liu2025}. Not all AGN exhibit extended emission, with those LAEs with the highest \lya line-fluxes preferring the single PSF component model. HSC-$r$ imaging reveals diverse morphologies, ranging from compact, continuum-bright sources to diffuse, continuum-faint systems with rest-frame equivalent widths exceeding 100\,\AA. Cross-matching with LOFAR radio data shows that radio counterparts are more common among the largest LANs.

Flux recovery comparisons confirm that PSF-based extractions systematically underestimate total Ly$\alpha$ flux. For HETDEX’s median seeing (FWHM $\simeq 1.8$\arcsec), extended-aperture fluxes are, on average, 30\% times larger than pipeline point-source values, implying that a significant fraction of Ly$\alpha$ light lies outside the core component.

The resulting \texttt{HETDEX LAN Catalog}, which accompanies this paper, provides positions, redshifts, and morphological and photometric parameters for the full \nlaegood\ LAEs. It can be found at \url{https://hetdex.org/data-results/} and in the electronic version of this paper. This catalog establishes the largest statistical census of extended Ly$\alpha$ emission to date, bridging the regimes of compact Ly$\alpha$ halos and luminous Ly$\alpha$ blobs across Cosmic Noon.  

\section*{Acknowledgments}

E.M.C. gratefully acknowledges the late Peter Erwin, developer of \texttt{pyimfit}\citep{Erwin2015}. His exceptionally efficient surface-brightness-modeling code made this work possible, enabling us to fit hundreds of thousands of objects over many iterative analyses.

HETDEX is led by the University of Texas at Austin McDonald Observatory and Department of Astronomy, with participation from the Ludwig-Maximilians-Universit\"at M\"unchen, Max-Planck-Institut f\"ur Extraterrestrische Physik (MPE), Leibniz-Institut f\"ur Astrophysik Potsdam (AIP), Texas A\&M University, The Pennsylvania State University, Institut f\"ur Astrophysik G\"ottingen, The University of Oxford, Max-Planck-Institut f\"ur Astrophysik (MPA), The University of Tokyo, and Missouri University of Science and Technology.  In addition to Institutional support, HETDEX is funded by the National Science Foundation (grant AST-0926815), the State of Texas, the US Air Force (AFRL FA9451-04-2-0355), and generous support from private individuals and foundations.

Observations for HETDEX were obtained with the Hobby-Eberly Telescope (HET), which is a joint project of the University of Texas at Austin, the Pennsylvania State University, Ludwig-Maximilians-Universit\"at M\"unchen, and Georg-August-Universit\"at G\"ottingen. The HET is named in honor of its principal benefactors, William P. Hobby and Robert E. Eberly. 

The Visible Integral-field Replicable Unit Spectrograph (VIRUS) was used for HETDEX observations. VIRUS is a joint project of the University of Texas at Austin,
Leibniz-Institut f\"ur Astrophysik Potsdam (AIP), Texas A\&M University
(TAMU), Max-Planck-Institut f\"ur Extraterrestrische Physik (MPE),
Ludwig-Maximilians-Universit\"at Muenchen, Pennsylvania State
University, Institut f\"ur Astrophysik G\"ottingen, University of Oxford,
and the Max-Planck-Institut f\"ur Astrophysik (MPA). In addition to
Institutional support, VIRUS was partially funded by the National
Science Foundation, the State of Texas, and generous support from
private individuals and foundations.

The authors acknowledge the Texas Advanced Computing Center (TACC) at The University of Texas at Austin for providing high-performance computing, visualization, and storage resources that have contributed to the research results reported within this paper. URL: http://www.tacc.utexas.edu

The Institute for Gravitation and the Cosmos is supported by the Eberly College of Science and the Office of the Senior Vice President for Research at the Pennsylvania State University. The Kavli IPMU is supported by World Premier International Research Center Initiative (WPI), MEXT, Japan. 

K.G. acknowledges support from NSF-2008793.

Software: This research was made possible by the open-source projects \texttt{hetdex-api}, \texttt{pyimfit/imfit} \citep{Erwin2015}, \texttt{astropy} \citep{astropy:2018}, \texttt{scipy} \citep{scipy}, \texttt{python} \citep{pythonref} and \texttt{numpy} \citep{harris2020array}.

\bibliography{hetdex}
\bibliographystyle{aasjournal}

\facilities{HET}

\end{document}